\documentclass[]{aa}
\usepackage{amsmath}
\usepackage{graphicx,color}
\usepackage{natbib}
\usepackage{microtype}
\usepackage{url}
\usepackage[varg]{txfonts}
\usepackage{threeparttable}
\usepackage{booktabs, makecell}
\usepackage{longtable}
\usepackage[utf8]{inputenc}
\usepackage{amsmath}
\usepackage{verbatim}

\def\gai{\textit{Gaia}\xspace}

\def\tes{\textit{TESS}\xspace}
\def\xmmn{\textit{XMM-Newton}\xspace}

\def\xro{\textit{XRONOS}\xspace}

\newcommand\vum{{V496~UMa}\xspace}

\newcommand\fergs{\ensuremath{\mathrm{erg}\,\mathrm{cm}^{-2}\,\mathrm{s}^{-1}}\xspace}

\newcommand\teff{\ensuremath{T_\mathrm{eff}}\xspace}

\begin{document}

\title{\xmmn and \tes observations of the highly variable polar V496 UMa\thanks{Based on observations obtained with \xmmn, an ESA science mission with instruments and contributions directly funded by ESA Member States and NASA}}
  
\author{Samet Ok \inst{1,2}
          \and
          Axel Schwope\inst{1}
          }
\institute{Leibniz-Institut für Astrophysik Potsdam (AIP), An der Sternwarte 16, 14482 Potsdam, Germany\\
\email{sok@aip.de}
\and
Department of Astronomy \& Space Sciences, Faculty of Science, University of Ege, 35100 Bornova, Izmir, Turkey
}

\date{Received December 2021; accepted May 2022}
 
\abstract
  {}
   {We aim to study the temporal and spectral behaviour of \vum from the optical to the X-ray regimes.}
   {We used archival \xmmn and \tes observations obtained in 2017 and 2019, respectively, to perform a spectral and timing analysis of the highly variable polar.}
   %
{The light curves of both satellites, \tes and \xmmn, reveal a double-humped pattern modulated with the periodicity of $91.058467 \pm 0.00001$ minutes. 
\vum displays a two-pole accretion geometry in the high accretion state. X-ray spectra from both regions are composed of thermal plasma radiation and soft blackbody components with almost identical temperatures and total accretion rate of $\dot{M}=1.4(8)\times10^{-11}$ M$_{\odot}$ yr$^{-1}$. 
The X-ray centers of the humps show longitudinal shifts of $-18\degr$ and $4\degr$ and $-172\degr$ and $-186\degr$ size around photometric phase zero, for the main hump and second hump, respectively. 
The long-term ZTF light curves reveal high and low accretions states. Low-state ZTF and SDSS photometric data are consistent with an $0.8$\,M$_{\odot}$ white dwarf at 10000 K and a main-sequence donor star with a spectral type of M5.0 at a \gai-determined distance of 758 pc.
}
{\vum is a very bright polar in X-rays when it is in the high state. Due to its unusual geometric structure, mass accretion onto the second accretion pole is interrupted occasionally. This discontinuous behavior does not follow a certain pattern in time and is observed so far only in the high state. The X-ray light curves display clear evidence for an accretion stream at the photometric phase of $\phi=$0.81, which does not show up in optical light curves. An accurate period was derived using the combined \tes and \xmmn data, which differs by 3.8 $\sigma$ from published results. 
}
\maketitle
%


\section{Introduction}
Cataclysmic variable stars (CVs) are accreting binary systems that transfer material via Roche lobe overflow from a late-type main-sequence star onto a white dwarf (WD). In the subgroup named the polars, the white dwarf's magnetic field strength is higher than $B> 10$\, MG and suppresses the formation of an accretion disk. Mass transfer occurs along magnetic field lines \citep{warner+95}. 

The material falling radially will go through a strong adiabatic shock at a certain distance above the surface which heats the infalling plasma to several 10s of keV, which subsequently cools via thermal plasma radiation at X-ray wavelengths and optical cyclotron radiation \citep{cropper+90}. The X-ray spectra of the polars are well described with a plasma that has a multi-temperature and variable density structure \citep[e.g.][]{cropper+90,fischer+01}. Polars are essential objects to understand the processes of mass transfer in the presence of a strong magnetic field, the generation of strong magnetic fields in close binaries, and the last steps of stellar evolution in the presence of a strong magnetic field.

\vum or MASTER OT J132104.04+560957.8 was firstly identified as a CV by \citet{yecheistov+12} when the system changed brightness by $\sim$2 mag from 18.41 mag to 16.43 mag within 30 minutes. Based on photometric observations, the system was tentatively identified as a non-eclipsing polar \citep{kato12}. It was confirmed as a polar based on its spectral properties by \cite{littlefield+15, littlefield+18}. Optical spectra showed strong hydrogen Balmer and strong He II 4686 \,\AA\ emission lines, and a non-stellar continuum that decreased blueward of 5000 \,\AA. Radial velocities and time-resolved photometry showed periodic variability at a period of 91 min, identified as the orbital period. Interestingly, the medium-resolution spectra obtained by \citet{littlefield+18} showed emission/absorption line-reversals at certain phases of the binary. Long-term Catalina Real-time Sky Survey \citep{drake+09} photometric observations revealed pronounced brightness variations during 2007 - 2012 \citep{littlefield+18}. The orbital light curve of the binary showed a double-peaked structure, with the amplitude of the second hump being strongly variable.

Precise and uninterrupted optical photometric data were available from \tes, the Transiting Exoplanet Survey Satellite \citep{ricker+14}. These data allow us to study such variations in much greater detail, which was a prime motivation for this study. These, along with unpublished \xmmn X-ray observations of the source, promise deeper insight into the accretion physics of this unusual polar.

We thus present an analysis of archival \xmmn, \tes and ZTF observations obtained between 2017 and 2021, respectively, when the system was in different accretion states. We aim to further explore the thermal, temporal, and geometric behavior of the object by analyzing the combined data set. In Section \ref{s:obs} we give the details of the observations. Section \ref{s:ana} contains our analysis and results. We conclude with a discussion of our results in Section \ref{s:dis}.

\begin{figure*}
\resizebox{\hsize}{!}{\includegraphics{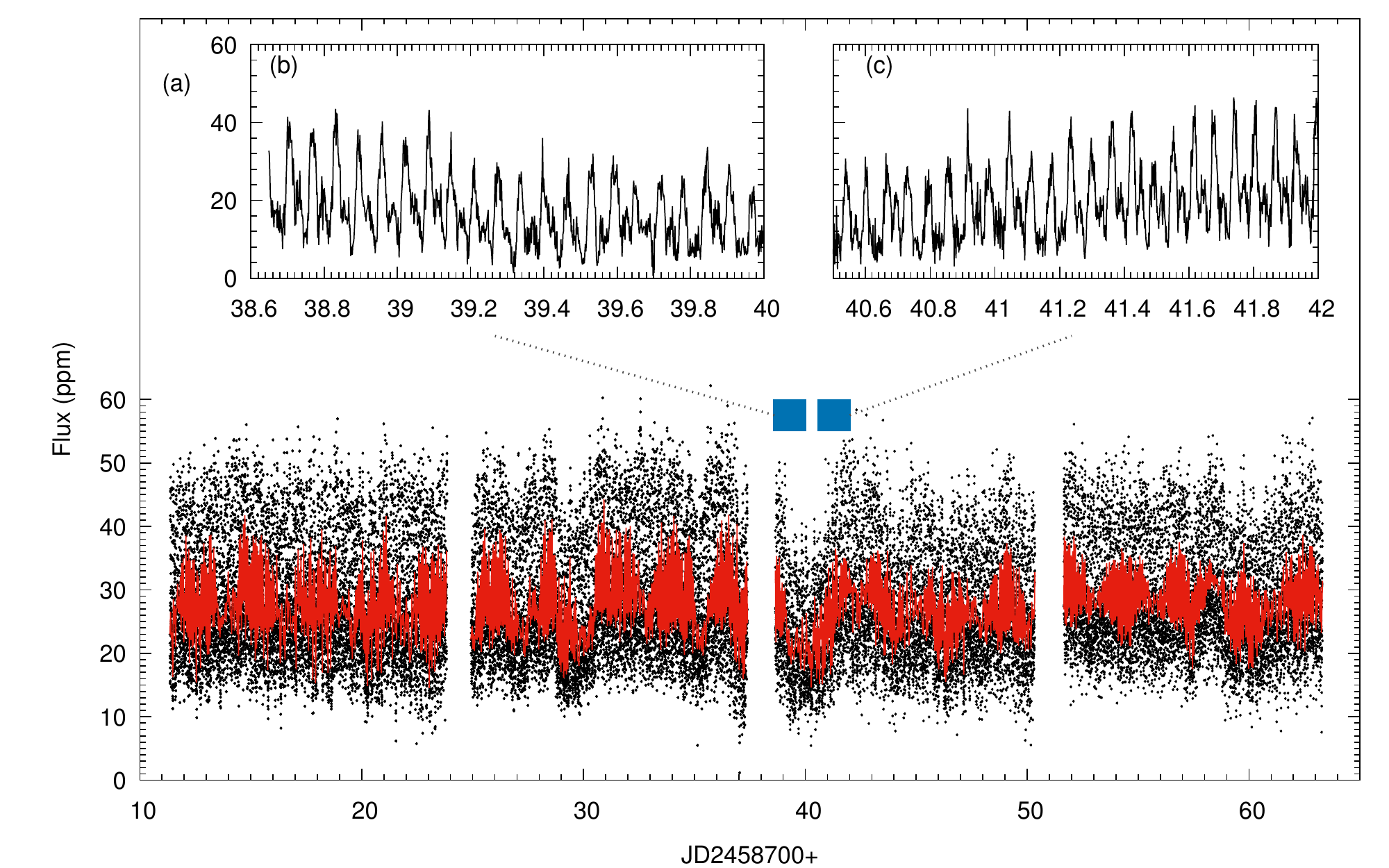}}
\caption{(a)\tes light curve of \vum obtained in 2019 with 2 min time resolution. The red line displays the binning time of 45 min. In the range of JD 2458739.1 and JD 2458741.4, the brightness of \vum displays increases and decreases suddenly. (b) shows the sudden decrease in brightness, while (c) focuses on the increase.
\label{f:lcori}}
\end{figure*}
\section{Observations \label{s:obs}}
\subsection{TESS Observations}

\tes is a space-based observatory providing high precision light curves for long uninterrupted time intervals. \tes is equipped with four identical refractive cameras with a combined field-of-view (FOV) of 24x96 degrees. 
Each of the 2k x 2k CCDs on the satellite has a scale of 21 arcseconds per pixel. 
\vum was observed in two sectors, and the photometric data for our target comes in two parts, named s0015-1001235444 and s0016-1001235444. The observations were made between Aug 15, 2019, and Sep 12, 2019, and were carried out in the high-cadence mode with a time resolution of two minutes. We use the Presearch Data Conditioning (PDC) light curves of this star produced by the TESS Science Processing Operations Centre (SPOC), which we downloaded from the Mikulski Archive for Space Telescopes\footnote{\protect\url{https://mast.stsci.edu}}. 

\begin{figure}
\resizebox{\hsize}{!}{\includegraphics[width=\columnwidth]{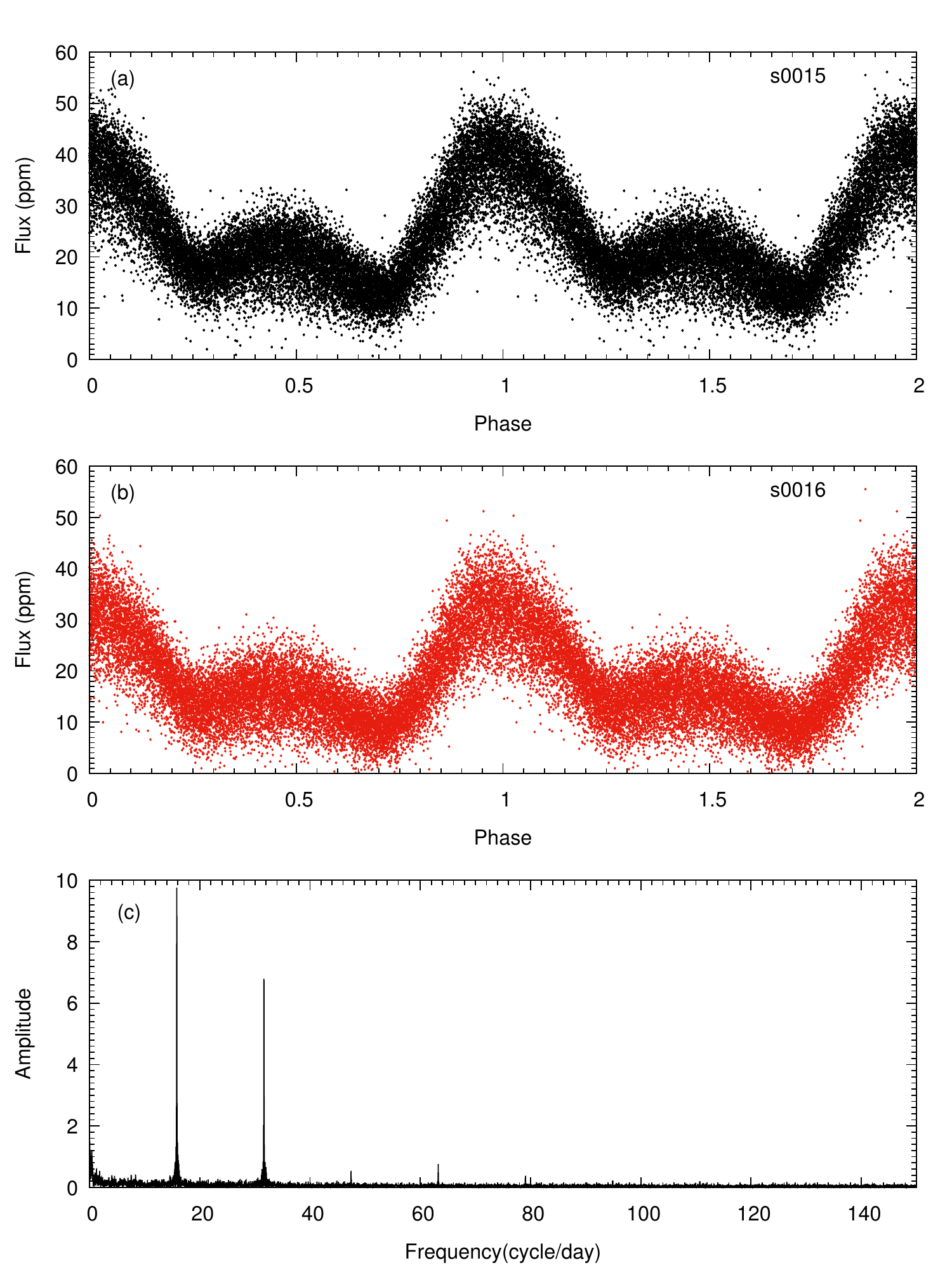}}
\caption{(a) and (b) show folded TESS light curve of \vum. Black and red dots indicate a continuous observation period of approximately 2 months. (c) Power spectrum of \vum. The highest amplitude frequency indicates the 91-minute period.
\label{f:lcfol}}
\end{figure}

We checked the apertures used to extract the light curves and found that two more sources fall into the same source extraction region in sector 16. Both objects are $\sim$10 arcsec away from our source. The coordinates of these objects obtained by \textit{Gaia} DR3 \citep{gaia+21} are RA1=13:21:06.03342, DEC1=+56:09:54.8807 (source id 1565396011799316352) and RA2=13:21:04.78039, DEC2=+56:10:09.4925 (source id 1565396007503248000), respectively. The \textit{G} brightness of these objects are 20.65(2) and 17.60(1).

We re-extracted the light curve by setting a new aperture for sector 16 using the python sub-package \textit{lightkurve} \citep{lcsoftware2018}. We used the same aperture value for sectors 16 and 15 and thus left the bright source with GaiaID 1565396007503248000 outside of the new aperture. The newly extracted light curve comprising the observations obtained in the two sectors is shown in Fig.~\ref{f:lcori}.

\subsection{Gaia Observations}
In \textit{Gaia} DR3 \citep{gaia+21}, \vum has the ID 1565395908719728384 and the brightness is 17.27$\pm0.03$, 17.31$\pm0.09$, and 16.86$\pm0.08$ in the \textit{G}, \textit{$G_{BP}$}, and \textit{$G_{RP}$} passbands, respectively. \textit{Gaia} measured the parallax of \vum as 1.307$\pm$0.060 mas.  We used the distance to the system of $758\pm33$\,pc, as determined by \cite{bailerjones+21}. 

\subsection{ZTF Observations}

We searched the Zwicky Transient Facility \citep{masci+19} database for data from \vum. \vum has coverage of three years (see Sec.~\ref{s:sed}) between March 2018 and June 2021. All data points were obtained with 30 sec exposure time, and in total, we found 566 data points in the \textit{g}, 544 points in the \textit{r} and 1556 points in the  \textit{i} passbands, respectively.

\subsection{X-ray Observation}

The \xmmn observations of \vum took place on 2017 March 12 (OBSID 0803500101), about two years before the \tes observations. They lasted 29 ks and thus covered 5.36 photometric cycles. The EPIC-pn and EPIC-mos cameras \citep{strueder+01,turner+01} were operated in full-frame mode with the thin filter. The Optical Monitor (OM) was set in the fast mode with an intrinsic time resolution of 0.5 s without using a filter. 
Data were reduced using the \xmmn SAS version 15.0.0. The EPIC-pn data were processed with the \textit{epchain} task to generate calibrated event lists. All times were corrected to the solar system barycentre using the \textit{barycen} task. Finally, the photon list was filtered to only include photons with energies between 0.2 keV and 10.0 keV for the EPIC-pn and EPIC-MOS instruments.
Our source extraction region was a circle centered on the source, and the best extraction radius was found with the \textit{eregionanalyse} task. The source extraction radii were 45 and 30 arcsec for EPIC-pn and EPIC-MOS instruments. An annulus was positioned around the object to extract the background photons. The radius of the inner ring of the annulus was determined as 50 arcsec, and the radii of the outer ring of the annulus were 110 arcsec.

Background-subtracted X-ray light curves were produced with the task \textit{epiclccorr} with 50\,s time bins. In addition, we calculated the Hardness Ratio (HR) obtained from different energy bands and used the HEASARC \textit{LCURVE} tool, a subpackage of \xro \citep{stella+angelini92}, and to display time-dependent light and hardness ratio curves (see Sec.~\ref{s:hr}).

One observation with the OM lasted 1200 s. A total of 10 such observations were made that also cover more than five orbital cycles. Between the exposures are short gaps without data that last 335 s. The OM data were reduced using the task \textit{omfchain} to produce background-subtracted light curves with 10 seconds time bins (see Sec.~\ref{s:omobs}).

\section{Analysis and results \label{s:ana}}
\subsection{Optical photometry obtained with \tes and \xmmn/OM}
\vum displays a highly variable behavior of the optical light curve. The light curve shape is characterized by two alternating humps with different brightness \citep[see Fig.~\ref{f:lcfol}, upper panel and][]{littlefield+18}.

To determine periodic changes in the light curve, we firstly computed a Fourier transformation using the Period04 software by \citet{lenz+bregel09}. We searched for significant peaks in the frequency interval from 0 $d^{-1}$ up to the Nyquist frequency of short cadence TESS data (360 $d^{-1}$). The power spectrum shows just one main period at f=15.81384(7) cycle d$^{-1}$ (P=91.0594(4) min) plus integer harmonics. To check for further possible periodicities, the leading term in the Fourier series computed by Period04 was subtracted from the original data (pre-whitening), and another power spectrum was computed. No further periodicity was found in the pre-whitened data. The uncertainty of the dominant frequency was then computed via the least square method in Period04 with the treatment of \citet{montgomery+winget99}. The power spectrum is displayed in the bottom panel of Fig.~\ref{f:lcfol}.

\begin{figure}[t]
\resizebox{\hsize}{!}{\includegraphics[width=\columnwidth]{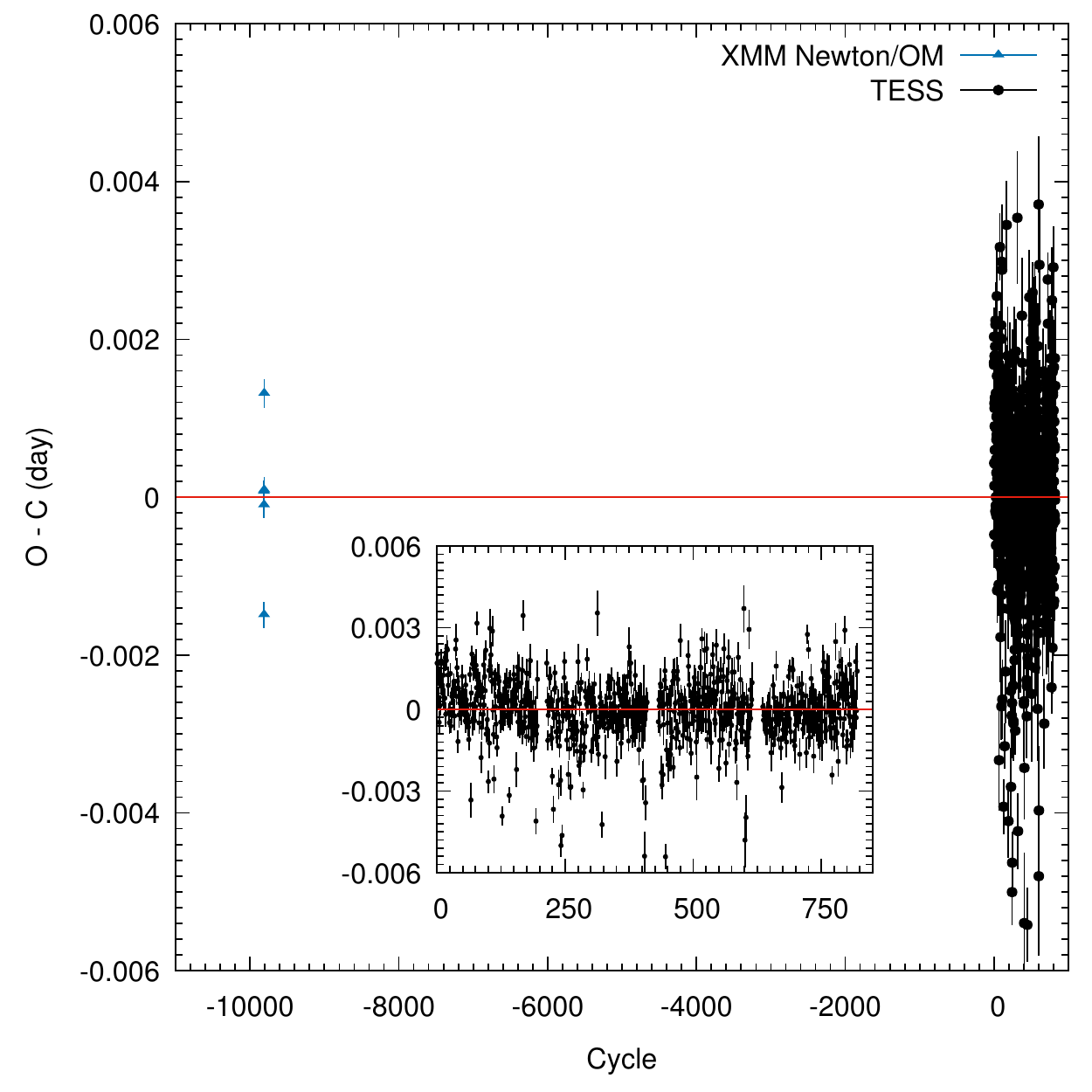}}
\caption{The $O-C$ variation of the main photometric hump in \vum with respect to the ephemeris of Eq.~\ref{e:eph} based on data from the OM onboard \xmmn and on \tes. The inset shows the \tes residuals only.
\label{f:omc}} 
\end{figure}

The main or first hump was always present in the lightcurves, and its arrival times were used to derive an improved optical ephemeris. We fit these bright humps with a Gaussian to obtain the times $T_{max}$ of their centers and their errors. The same process was applied to the \xmmn Optical Monitor light curve (see Fig.~\ref{f:xmmlc}). Using both \xmmn/OM and \tes, a weighted linear regression between the times $T_{max}$ and the cycle numbers were used to derive

\begin{equation}
BJD(T_{0,max}) = 2458711.45213(2) + 0.0632350467(7) \times E
\label{e:eph}
\end{equation}

where the numbers in parenthesis give the uncertainties in the last digits.
The residuals of this linear fit are shown in Fig.~\ref{f:omc}. The error of the period obtained from the Fourier transformation is larger than that from the linear regression. We, therefore, accept the results from the linear regression as our final timing solution.

The ephemeris given in Eq.~\ref{e:eph} improves the previous period determination of \citet{littlefield+18} considerably. While the period they derive is $0\fd06323520(4)$, our new value has a factor $\sim$60 smaller error. This improvement was possible using the quality of the \tes data and the time-base of the combined \tes and \xmmn/OM data. 

In the following, all phases refer to the ephemeris given in Eq.~\ref{e:eph}. 
We note that the previous period value deviates by $3.8\sigma$ (of the old larger error) from our new value. The time difference between the two ephemerides is 2426.64783 days which corresponds to 38375.046 cycles according to our new period. Hence, our phase zero is later than the old value by 0.046 phase units.

\begin{figure}
\resizebox{\hsize}{!}{\includegraphics[width=\columnwidth]{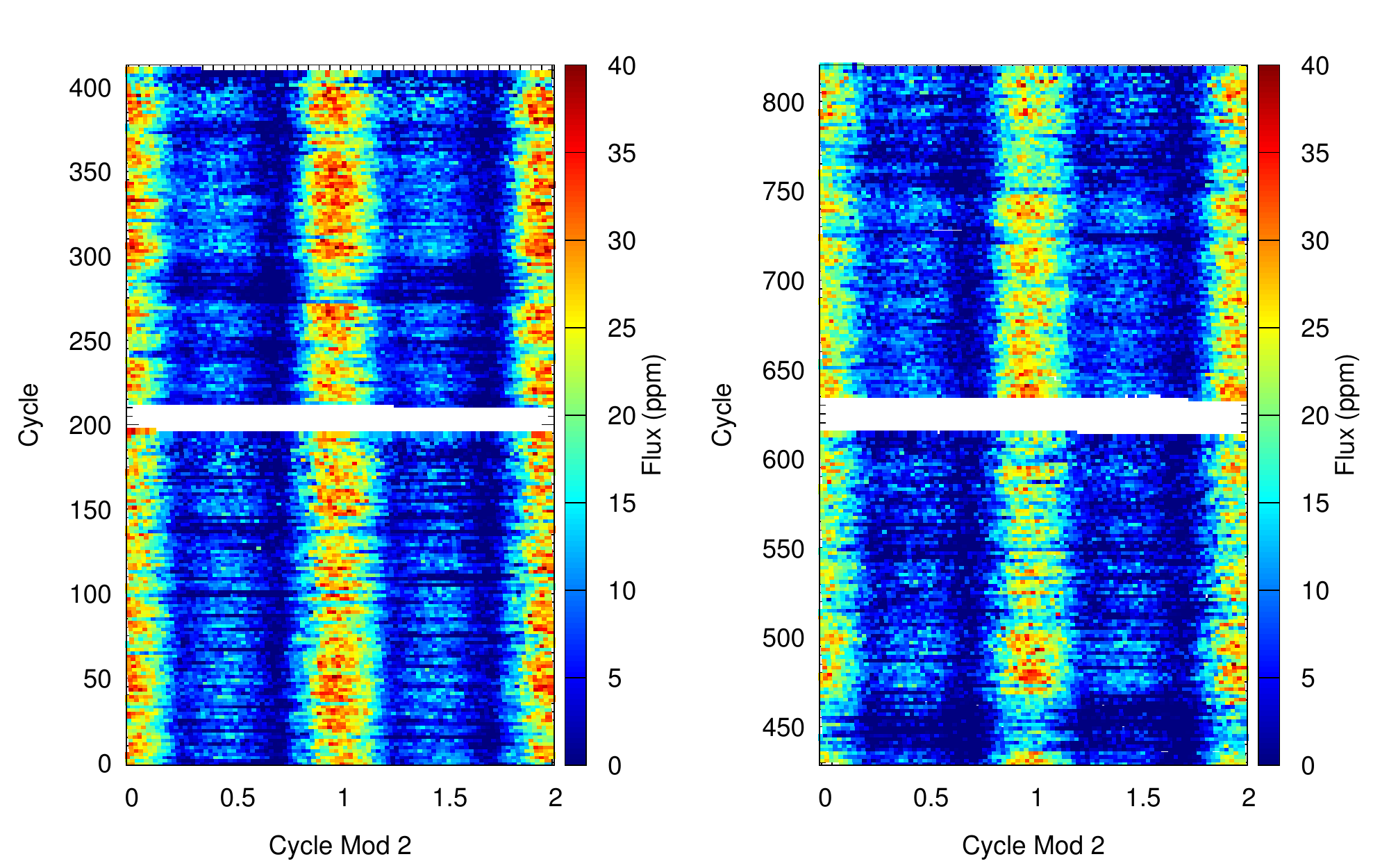}}
      \caption{Cycle - Cycle Mod 2 diagram of \vum for all high-cadence \tes data. Two consecutive cycles are drawn consecutively from the beginning. The color bar shows the flux density. The left panel shows data obtained in sector 15, the right panel is obtained in sector 16. All data were folded using the ephemeris given in Eq.~\ref{e:eph}. The wide white gaps in both graphs show where \tes did not receive data.
\label{f:tlc2d}
} 
\end{figure}

We use our new ephemeris to create an epoch-phase diagram that shows all the data obtained with \tes in a compact format in Fig.~\ref{f:tlc2d}. TESS observations cover about 800 photometric cycles. The long-term light curve of \vum shows the overall stability of the light curve pattern on the one hand, with some marked variability on the other hand. There are changes in the overall brightness affecting both humps, and there are occasions when apparently only the second hump was found to be strongly variable. This behavior was already reported by \citet{littlefield+18} based on ground-based time series. The continuous \tes data offer new insight into the frequency and the duration of such changes.

Fig.~\ref{f:tlc2d} shows a considerable decrease in the overall flux happening between cycles 268 - 300 and 430 - 470. The overall light curve shape with two humps remains more or less the same, although there is some suggestion that the second hump would be affected more strongly than the first during those overall brightness reductions. To further investigate this behavior, we extracted the brightness of the two humps per cycle.
For this, we identified three phase intervals from the phase folded light curve, 0.92 - 1.07 for the first hump, 0.36 - 0.55 for the second hump, and 0.68 - 0.72 for the faintest part of the light curve, which was used as the base flux or background and subtracted from the integrated flux in the two humps.

\begin{figure}
\resizebox{\hsize}{!}{\includegraphics[width=\columnwidth]{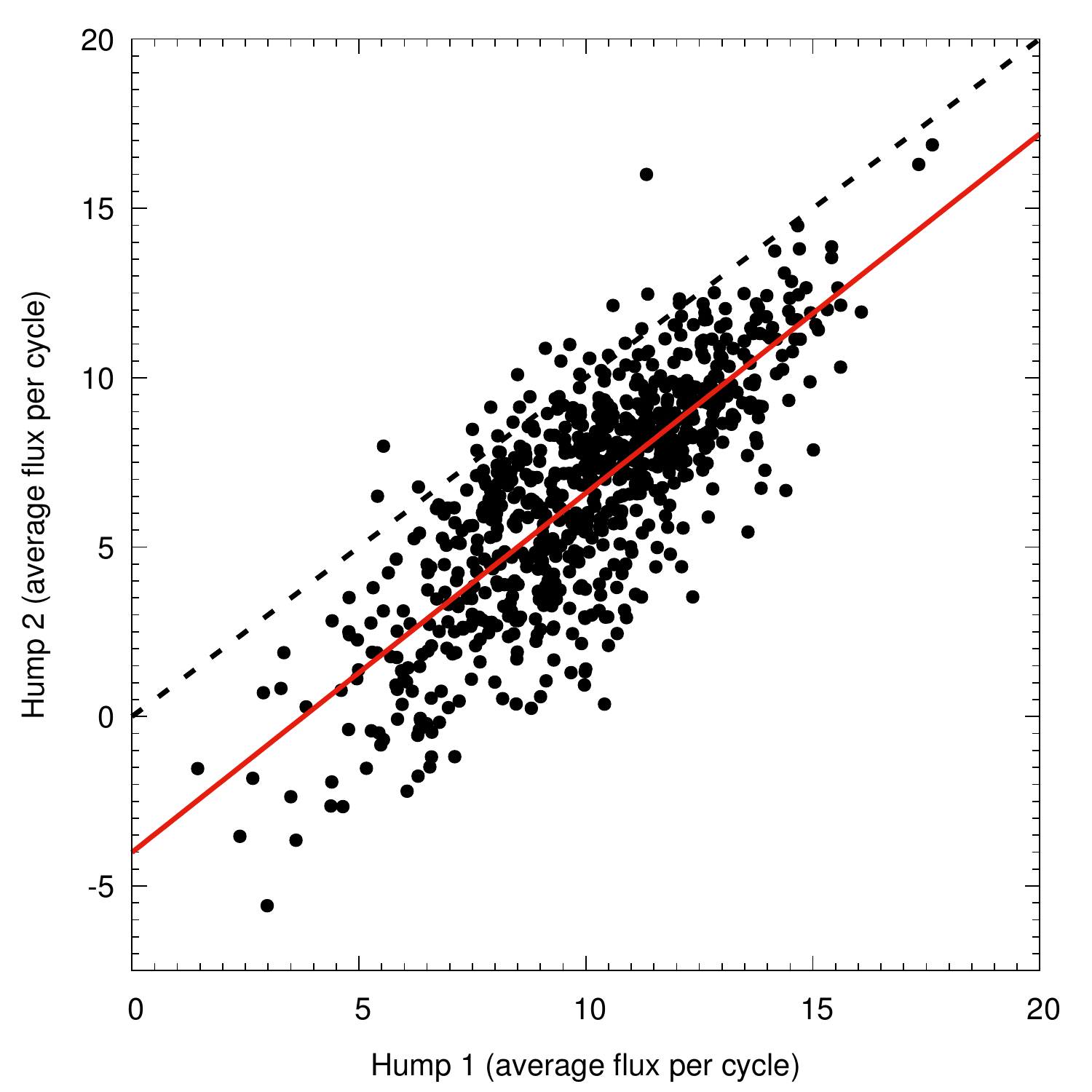}}
      \caption{Hump 1 vs. Hump 2 diagram. Average fluxes in the phase range of 0.92 - 1.07 (Hump 1) and 0.36 - 0.55 (Hump 2) in each cycle were calculated. The average base corresponding between 0.68 - 0.74 phase in each cycle was subtracted from the related humps. The red line shows the linear fit corresponding to the humps. The dashed line shows where the flux in both humps are equal.
\label{f:hump1vs2}
} 
\end{figure}

Fig.~\ref{f:hump1vs2} shows the variation of both related humps in the same cycle, which displays a linear relation. Also, this linear relationship is parallel to the line where the flux of both humps is equal. The first hump always has more flux than the second hump and always has positive flux, while the second hump is consistent with zero flux at the lowest levels. The average flux ratio between hump one and hump two is 1.89.

The linear relation nevertheless shows a large scatter, implying that the distribution of matter onto the one or the other pole that might be associated with the two light-curve humps (if a two-pole scenario applies, see below for a discussion) instantaneously shows significant changes.

Since the system does not show an eclipse, it is difficult, if not impossible, from optical photometry alone to decide whether mass accretion occurs via a single or two poles. Both geometries can produce a double-humped light curve. For instance, in a one-pole geometry, two humps may occur through stream obscuration and cyclotron beaming, while in a two-pole geometry, two humps occur through successive self-eclipses of the two poles.

X-ray observations play a crucial role in understanding which of these two scenarios occur. The hydrogen column density in the direction of the object ($N_{H}$) and hardness ratio (\textit{HR}) behavior in the dips and bright humps can facilitate understanding of the proper geometry of the \vum (see next section).

\subsection{\xmmn observations}
\subsubsection{X-ray photometry \label{s:omobs}}

\begin{figure}
   \centering
   \includegraphics[width=\columnwidth]{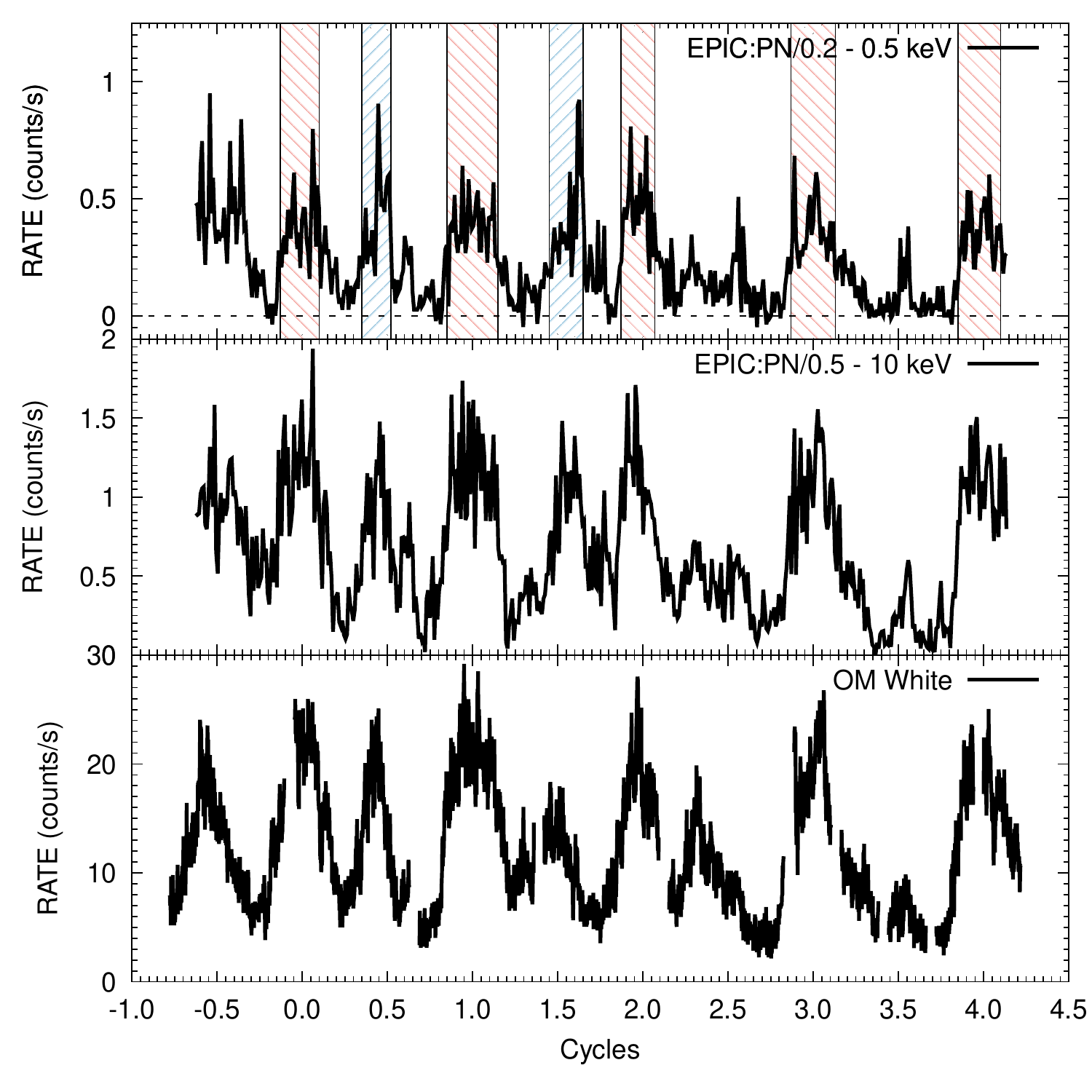}
      \caption{X-ray and optical light curves of \vum obtained \xmmn. The top two panels show the light curves obtained in the $0.2 - 0.5$ keV and $0.5 - 10$ keV ranges with time bins of 50 sec. The OM data in the lower panel have time bins of 10 s. In the upper panel, red (hump 1) and blue (hump 2) filled shaded regions identify the phase intervals used to extract the X-ray spectra of both humps.} 
\label{f:xmmlc}
  \end{figure}

In Fig.~\ref{f:xmmlc} the light curves obtained with the \xmmn observatory at X-ray and optical/UV wavelengths are shown in the original time sequence. 
Similar to \tes in the optical, the light curves at X-ray wavelengths show a double-humped shape. The first hump is mainly constant in brightness, while the second hump is highly variable and unstable. These observations covered almost five complete photometric cycles, beginning with a second hump. The first hump is a factor 1.3 brighter than the second hump during the first two photometric cycles ($0.2-10$\,keV). The first and the second humps were separated by about 0.53 phase units in these cycles. During the last two cycles, the second hump was much fainter. In the phase interval 0.8 - 1.2 (hump 1) and 0.3 - 0.7 (hump 2) of the last two cycles, the mean flux ratios increased to 2.3 and 2.6.
The simultaneous OM observations show that weakening of the second hump in the optical to UV region starts already (after JD2458091.1) in the second full photometric cycle, i.e.~one cycle earlier than observed at X-ray wavelength.

\begin{figure}
   \centering
   \includegraphics[width=\columnwidth]{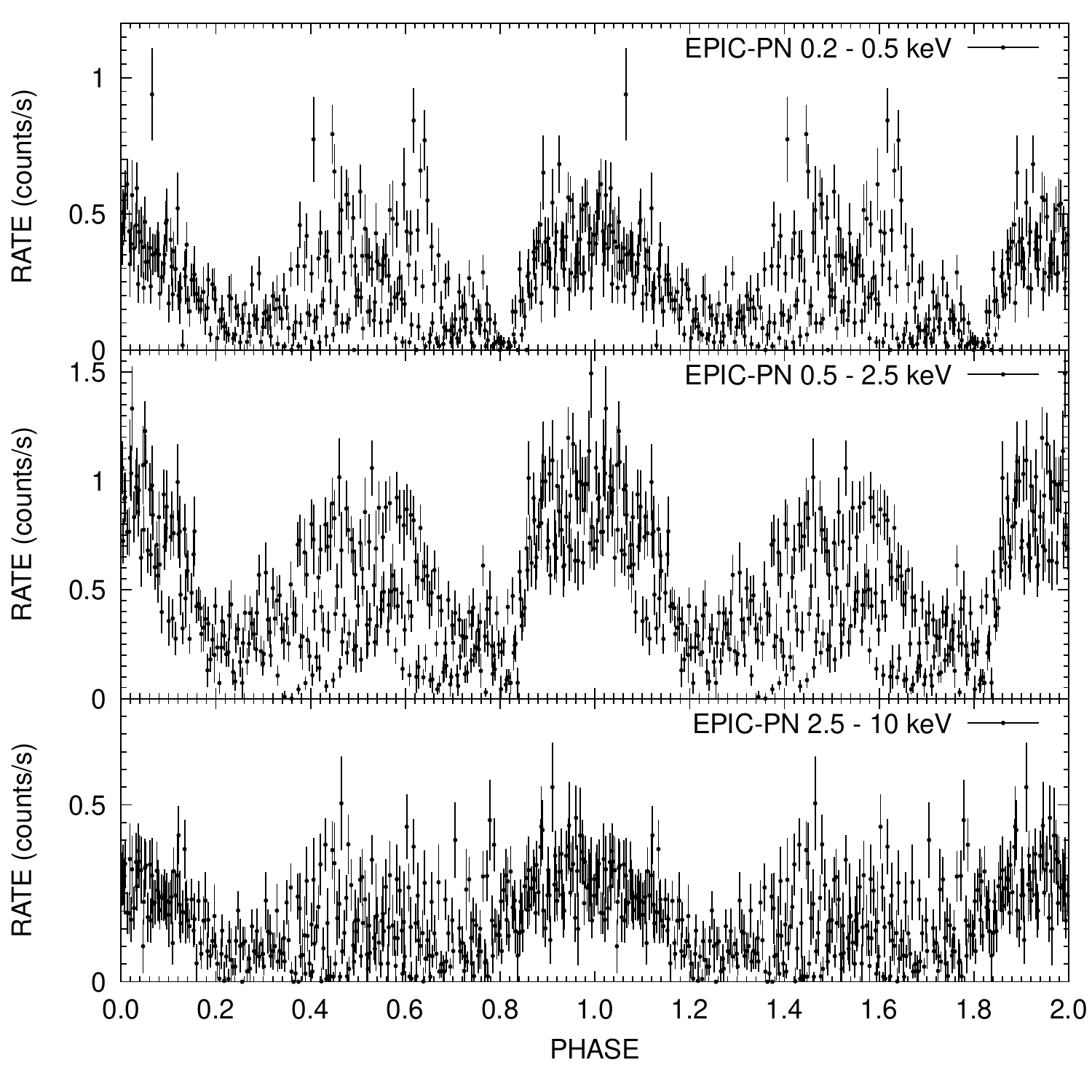}
      \caption{Energy resolved X-ray light curves of \vum. They were folded according to the Eq.~\ref{e:eph}. The light curves were extracted using time bins of 50 seconds.} 
\label{f:xmmlcclr}
  \end{figure}

We also generated energy-resolved, phase-folded X-ray light curves in the $0.2 - 0.5$ keV, $0.5 - 2.5$ keV and $2.5 - 10$ keV energy bands. These are displayed in Fig.~\ref{f:xmmlcclr}. It shows the main hump as a rather stable feature in all bands with always the same width and similar brightness. A dip is seen in the soft energy band (0.2 - 0.5 keV) at phase $\phi$ = 0.81 with zero rates in all covered binary cycles. The width of the dip is about $\Delta\phi$ = 0.037 (200 sec). The dip is definitely not due to the donor star's eclipse of the accreting white dwarf. Such a binary eclipse would be visible at all energies. Instead, the feature is more likely a stream dip, where the X-rays from the accretion region on the white dwarf are passing through the accretion stream that was raised out of the orbital plane. Its origin by photoelectric absorption in the stream naturally explains its energy dependence.

\subsubsection{X-ray spectroscopy \label{s:hr}}

The X-ray spectra obtained with \xmmn were analyzed with version 12.8.2 of the XSPEC package \citep{arnaud+96,dorman+01}. Spectral fitting was performed for EPIC-pn and EPIC-MOS in the energy range of $0.2 - 10$ keV. Mean spectra were extracted for the first and second hump for the time intervals shown in Fig.~\ref{f:xmmlc}.

The background noise was high (\textit{RATE} >1) during the first 30 minutes of the observation (until the first minimum in the light curve), which was therefore excluded from our analysis. We used MOS1 and MOS2 data to analyze the mean spectra from hump 1 and hump 2 but not for time-resolved spectroscopy due to their low signal-to-noise ratio. 

The spectrum of the first hump was obtained from the first five stable peaks seen on the light curve (see Fig.~\ref{f:xmmlc}). The spectrum of the second hump was acquired only for the first two full occurrences due to the weakness of this feature during the last two photometric cycles. Both spectra were grouped with 25 counts per bin. In the spectral analysis, we used the ${\mathrm \chi}^{2}$ statistics for model optimization.

\begin{figure}
   \includegraphics[width=\columnwidth]{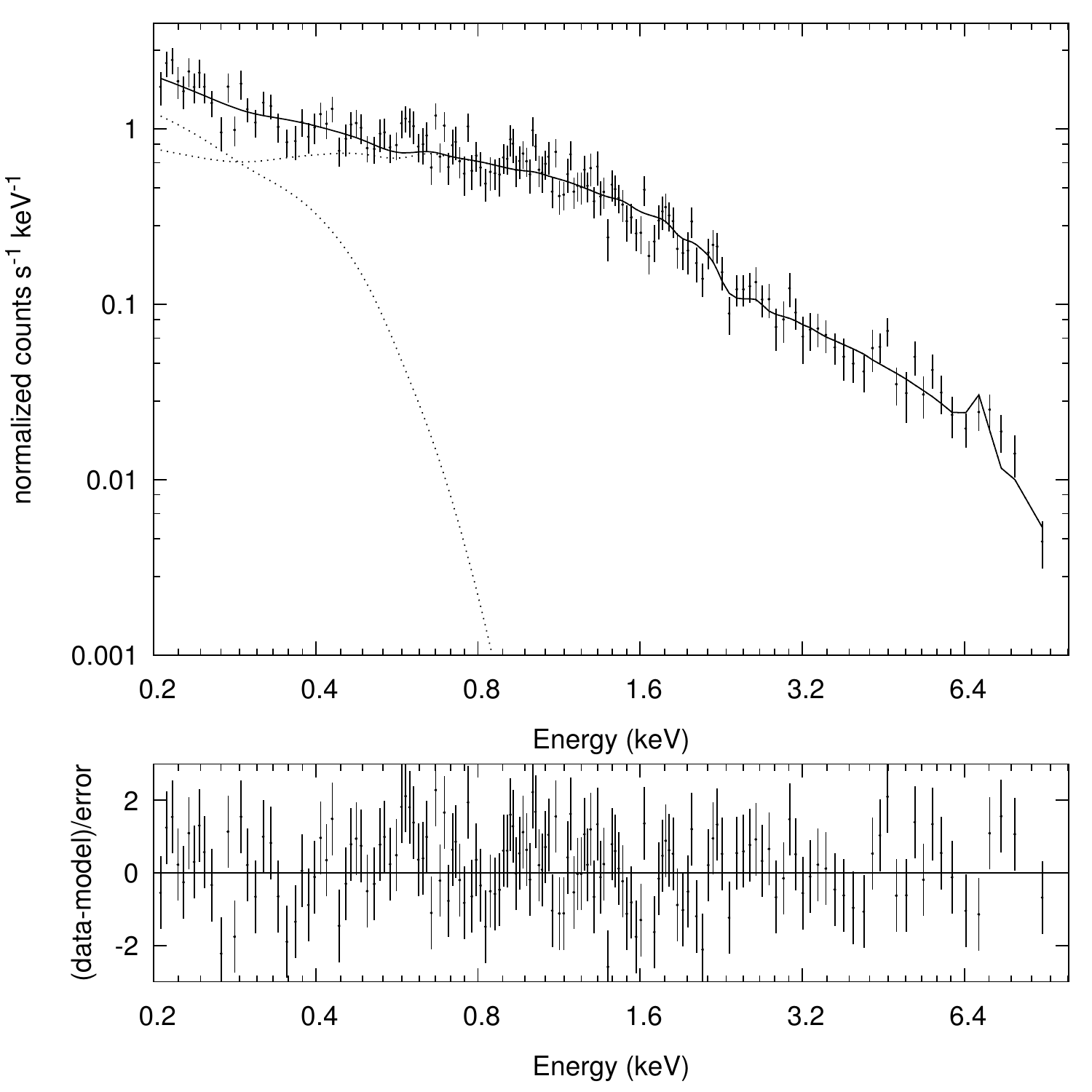}
      \caption{X-ray spectrum of  \vum. The spectrum has been extracted from the first bright humps and grouped with 25 counts per bin. We used  {\tt TBABS(BB+MEKAL)} spectral models. The reduced chi-square value is 1.15. } 
      \label{f:hump1pha}
\end{figure}

\begin{figure}
   \includegraphics[width=\columnwidth]{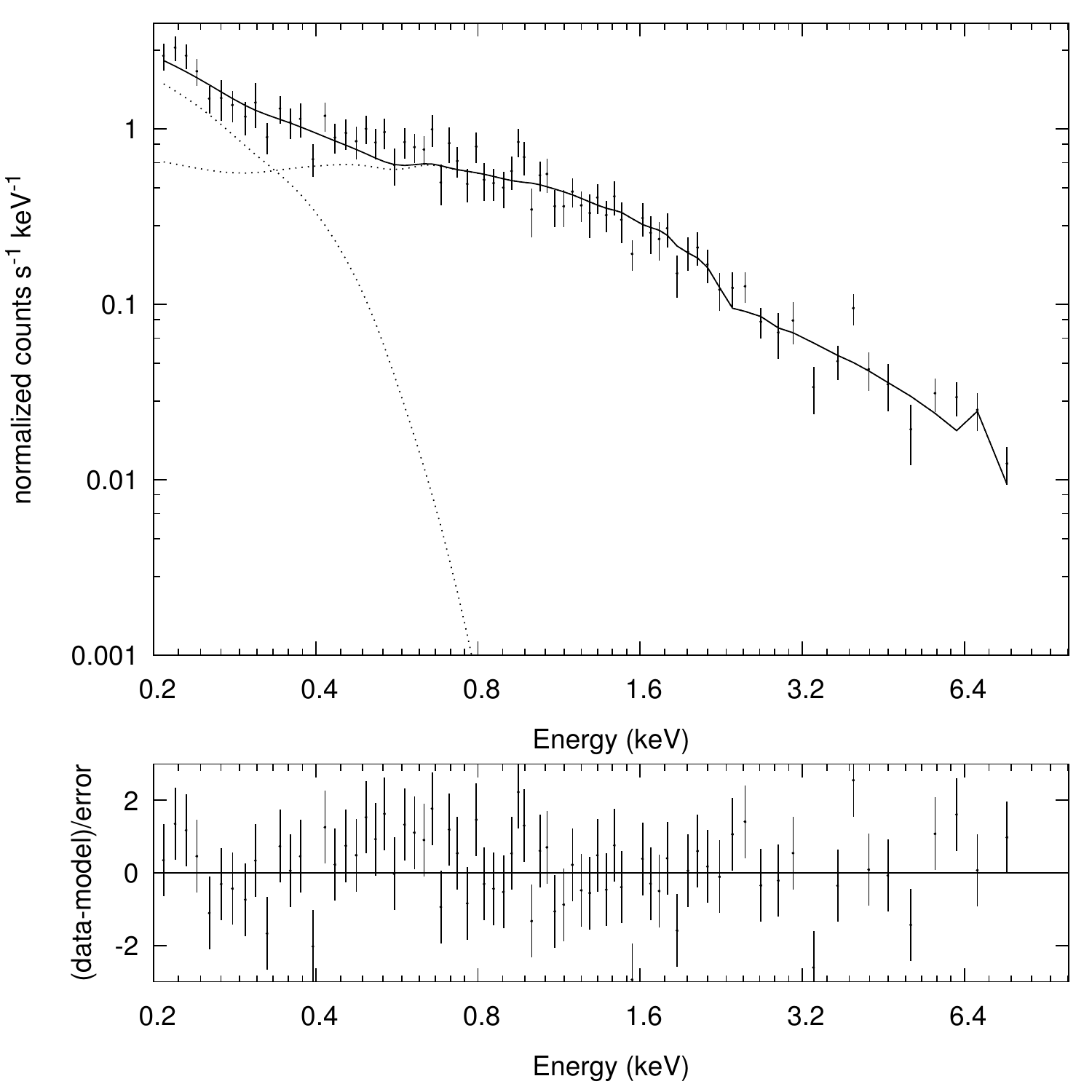}
      \caption{X-ray spectrum of \vum. The spectrum has been extracted from the first two peaks of the second hump, which can be seen in the light curve in Fig.~\ref{f:xmmlc} and grouped with 25 counts per bin. We used the same models that were applied to the first bright hump. The reduced chi-square value is 1.20.} 
      \label{f:hump2pha}
\end{figure}

\bgroup
\def\arraystretch{1.25}
\begin{table}
\centering
\caption{Spectral parameters from humps: Spectral fit parameters, their uncertainties, fit statistics, and model bolometric fluxes.}
\label{t:1}
\resizebox{\columnwidth}{!}{
\begin{tabular}{lccc}
\hline       
\hline                     
\multicolumn{3}{l} {Model: {\tt TBABS(BBODY+MEKAL)}}                                \\
\hline
                         &  First Hump               &  Second Hump                         \\
     \hline
Parameters               &  EPIC-PN                   &  EPIC-PN                           \\
\hline
   
$kT_{bb}$(eV)           & $51^{+10}_{-9}$              & $43^{+10}_{-8}$                   \\
$kT_{mekal}$(keV)       & $10.7^{+2.9}_{-2.0}$          & $9.8^{+4.7}_{-2.5}$            \\

$\chi^{2}$(d.o.f)       & 1.15(174/151)                 & 1.20(84/70)                        \\
\hline
Unabsorbed Fluxes\\
\hline
$F_{0.5-2.5} (\fergs)(10^{-12})$  & $1.50^{+0.05}_{-0.05}$ & $1.28^{+0.06}_{-0.06}$          \\
$F_{2.5-10}  (\fergs)(10^{-12})$  & $2.6^{+0.2}_{-0.2}$ & $2.1^{+0.1}_{-0.1}$          \\
$F_{bol}     (\fergs)(10^{-12})$  & $6.7^{+0.4}_{-0.4}$   & $7.3^{+0.4}_{-0.4} $                      \\
\hline
$F_{bb}     (\fergs)(10^{-12})$  & $1.2^{+0.8}_{-0.4}$    & $2.7^{+0.4}_{-0.3} $                       \\
$F_{mekal} (\fergs)(10^{-12})$  & $5.4^{+0.4}_{-0.4}$     & $4.5^{+0.5}_{-0.5}$                       \\
\hline
$L_{x}     (\fergs)(10^{32})$    & $2.3(3)$              & $2.5(4) $                       \\                   
\hline
\end{tabular}
}
\end{table} 
\egroup

\begin{figure}[h]
   \centering     
   \includegraphics[width=\columnwidth]{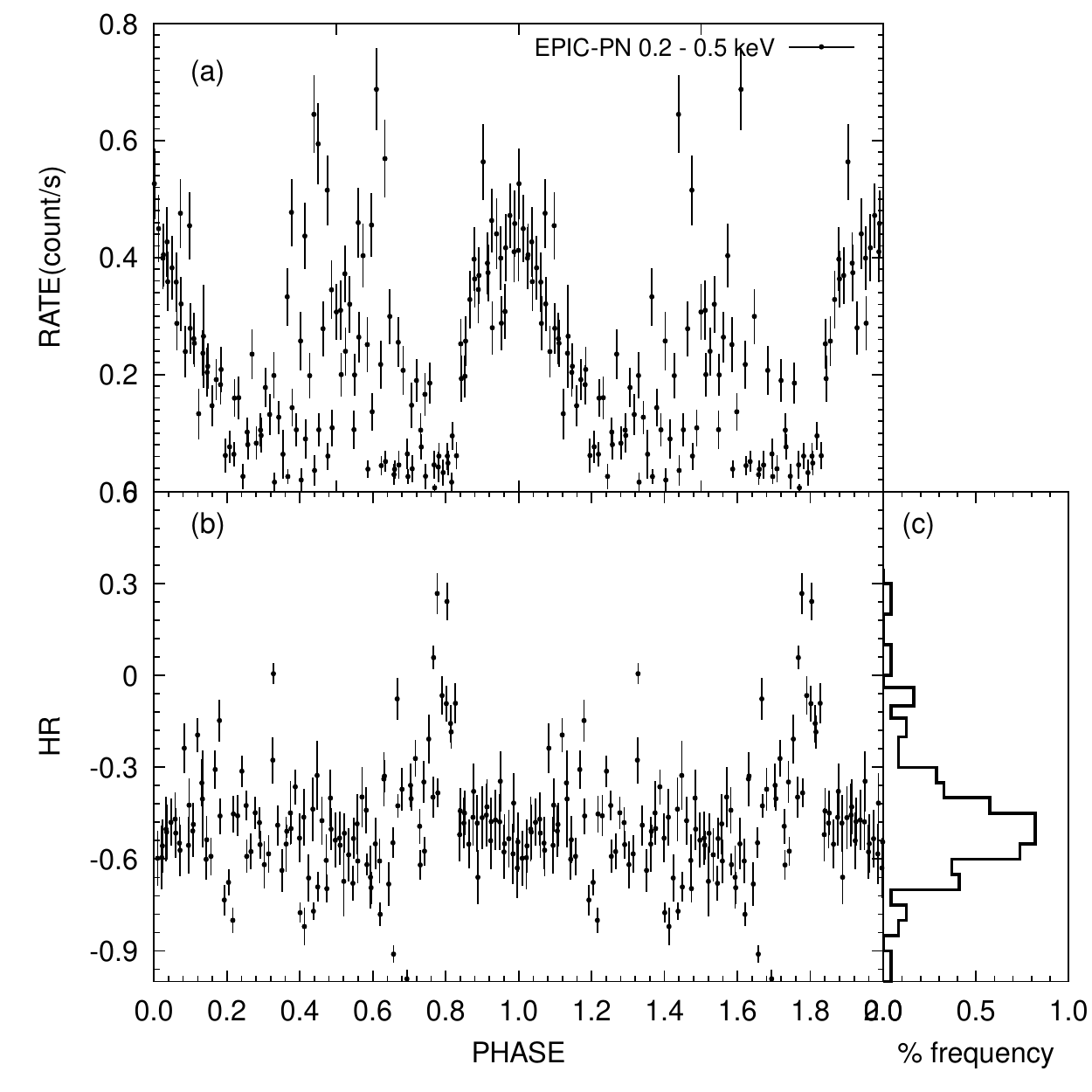}
      \caption{(a) Folded X-ray light curve of \vum (0.2 - 0.5 keV). (b) Hardness Ratio variation of \vum which corresponds to energy range between 0.5 - 2.5 keV and 2.5 - 10 keV. (c) shows the frequency of HR along the folded X-ray light curve.}
      \label{f:lcnhhr}
\end{figure}

Following \citet{kuulkers+10} and \citet{mukai17} (and references therein), the bright phase spectra of polars can be described with optically thin thermal plasma emission from the post-shock plasma, sometimes with an additional blackbody component from the accretion-heated white dwarf atmosphere, and we follow this approach here as well. 

We started to fit the first hump data using a single {\tt  MEKAL} plasma model \citep{mewe+85,liedahl+95} with the metal abundances frozen at solar values \citep{wilms+00}. \vum is located at a very high galactic latitude and is a relatively distant object compared to the distribution of galactic hydrogen. We decided to include an absorption component to the {\tt  MEKAL} and to fix the absorbing column at the galactic value, $N_{\rm H,gal}$ =1.23$\times10^{20}$ atoms cm$^{-2}$ \citep{bekhti+16}.  The combined simple model {\tt TBABS*MEKAL} gave a bad fit to the data with a reduced $\chi^{2}=2.14$ (for 328/153 d.o.f). The fit left an excess of photons in the soft spectral region, $<0.5$\,keV.

We then added a single blackbody model ({\tt BB}) so that our new model is {\tt TBABS(BB+MEKAL)}. The fit yielded a chi-square value of 174.35 for 151 degree of freedom, $\chi_\nu^2 = 1.15$. The null hypothesis probability of rejecting the model was 9\%. We thus accepted this model for the first hump (see Fig.~\ref{f:hump1pha}).

A similar process was applied for the spectrum of the second hump. For this, we also started with a {\tt TBABS*MEKAL} model. This initial fit yields a chi-square of 208.72 for 72 degrees of freedom ($\chi^{2}$=2.89) and was thus rejected. The fit left an excess of photons mainly at soft energies below 0.8 keV. Then, we added a {\tt BB} model to describe this energy range better. The fit was improved substantially with a chi-square value of 83.93 and 70 degrees of freedom. Our reduced $\chi^{2}$ value decreased to $\chi^{2}=1.19$. The probability of rejecting the null hypothesis is 12\%, and we thus accept the emission model with a blackbody and a thermal component. The spectrum, the best-fit model and the residuals are shown in Fig.~\ref{f:hump2pha}.

The best-fit model parameters are summarized in Table \ref{t:1}. Their errors were calculated with the \textit{error} and \textit{steppar} commands in XSPEC for 90\% confidence levels (the delta fit statistic of 2.706). Uncertainties on fluxes for all models were calculated using the \textit{cflux} convolution component. Our spectra contain too few photons to derive any useful constraint on the existence and the parameters of possible Fe-line complex between 6.4 and 7 keV. 

The spectral shapes and temperatures of both humps are very similar and compatible with each other within the error limits. The most notable difference was in the bolometric fluxes of the blackbodies. The bolometric flux of the second hump is a factor 2 than that of the first hump. This is due mainly to the lower temperature found for the blackbody from the second hump, which places a larger part of the spectrum in the unobserved EUV region of the spectrum. Therefore, the reliability of this calculation is debatable because we only see the extension of this blackbody component in the X-ray range. On the other hand, we emphasize that in the 0.2 - 0.5 keV soft energy range, where the second hump is prominent, it has higher count numbers than the first hump (see Fig.~\ref{f:xmmlcclr}, blue-shaded regions). 

For the distance of 758 pc, the soft blackbody emission of the first hump originates from an assumed circular region with a radius $690^{+620}_{-300}$ km, that of the second hump with a radius $1450^{+1350}_{-430}$ km. These radii correspond to fractional areas of an assumed WD with a radius of 8000\,km of 0.8\% and 0.2, respectively.

The second region is possibly larger than the first, but this is uncertain and we do not speculate about possible reasons. However, taking at face value, the implied sizes of the emission region are larger than those found e.g.~in the prototypical object AM Herculis \citep[about 100 km,][]{schwope+20} but of similar size as in the textbook object EF Eri \citep[$<$ 570 km][]{beuermann+87} and thus approximately as expected given the large uncertainties of the derived values and on the heating processes in this object. For a discussion of sizes of emission regions, see e.g.~\cite{hameury+king88}.

We also calculated hardness ratios $ HR= ((H-S)/(H+S))$ from the counts in the two energy bands ($S=0.5 - 2.5$\,keV and $H= 2.5 - 10$\,keV) from background-corrected light curves with 200 s time bins. Figure~\ref{f:lcnhhr} shows the change of \textit{HR} for the whole X-ray observation. The only significant and repetitive feature in the hardness ratio curve is a substantial increase in the dip before the first bright hump at $\phi$=0.81(1) (see Fig.~\ref{f:lcnhhr}). The dip, therefore, is naturally explained by photoelectric absorption in the intervening accretion stream lifted out of the orbital plane. To estimate the amount of absorption in the dip ($\phi$=0.81), we ran XSPEC models using the model parameter of hump 1, adding an extra {\tt TBABS} component. We found a dip $N_{H}$ of order $1.6\times10^{21}$ atoms cm$^{-2}$.
 
Apart from this, no increase or significant variation in the hardness was observed during the first and second humps. In particular, we note that the hardness stays the same independent of the pronounced brightness variations of the second hump. We regard this as evidence that accretion occurs at two different accretion zones on the white dwarf surface. In other words, the system is a two-pole accretor. If there were just one accreting pole, the brightness of the two humps would change coincidentally. The occasional reduction of the brightness of the second hump would be due to absorption in an accretion curtain. Such absorption events would have been evident through an increase in the hardness ratio, which is not observed.

\begin{figure}[t]
   \includegraphics[width=\columnwidth]{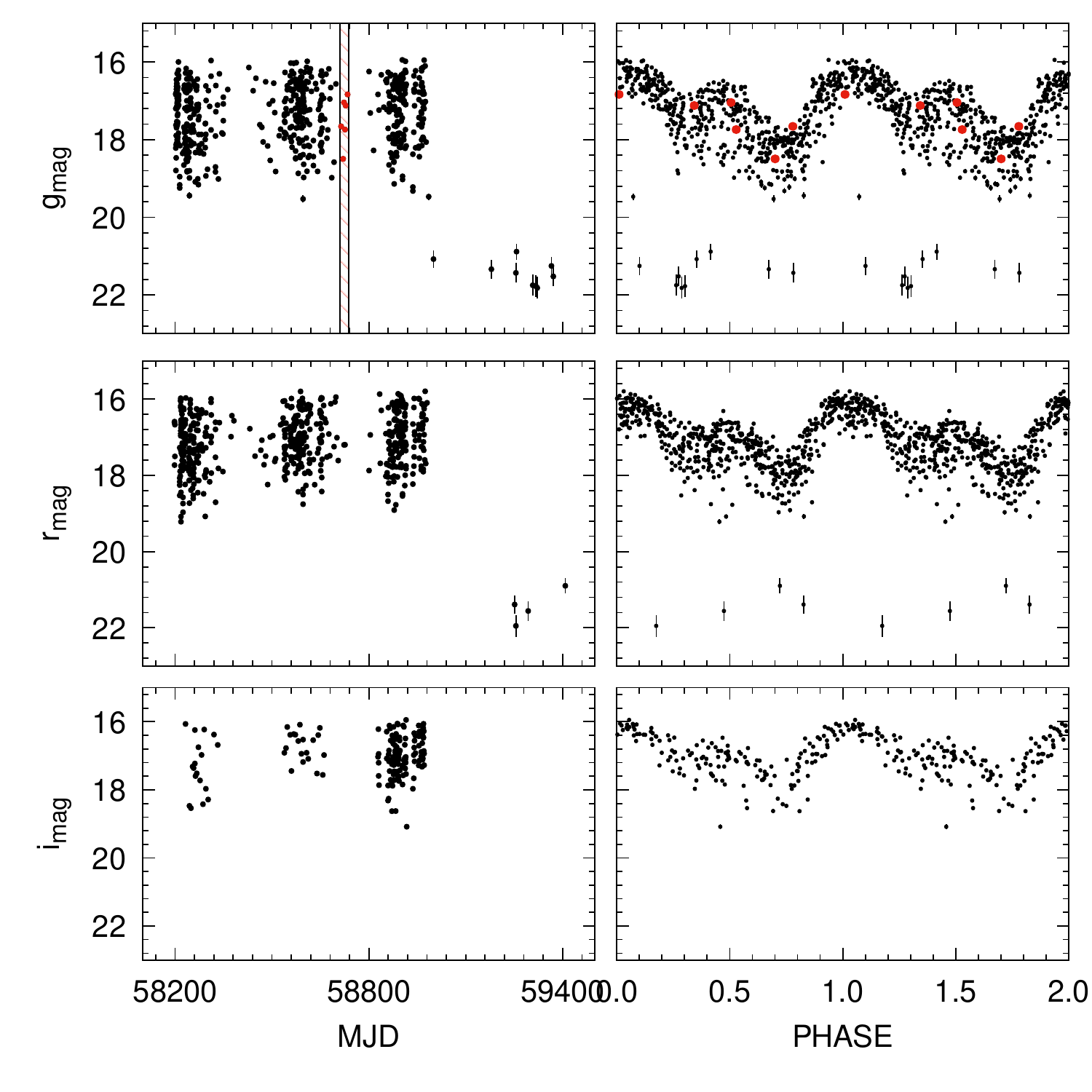}
      \caption{The three figures from top to bottom on the left show the ZTF observations in time series with different passbands. The figures on the right side show that folded light curves of these time series according to Eq. 1. The interval (dashed red box) in the upper left figure shows the time when \tes observations were made. The ZTF observations at this time are given by the red dots. The phase folded figure on the right shows the location of these dots. } 
      \label{f:ztf}
\end{figure}

\begin{figure}[t]
   \centering
   \includegraphics[width=\columnwidth]{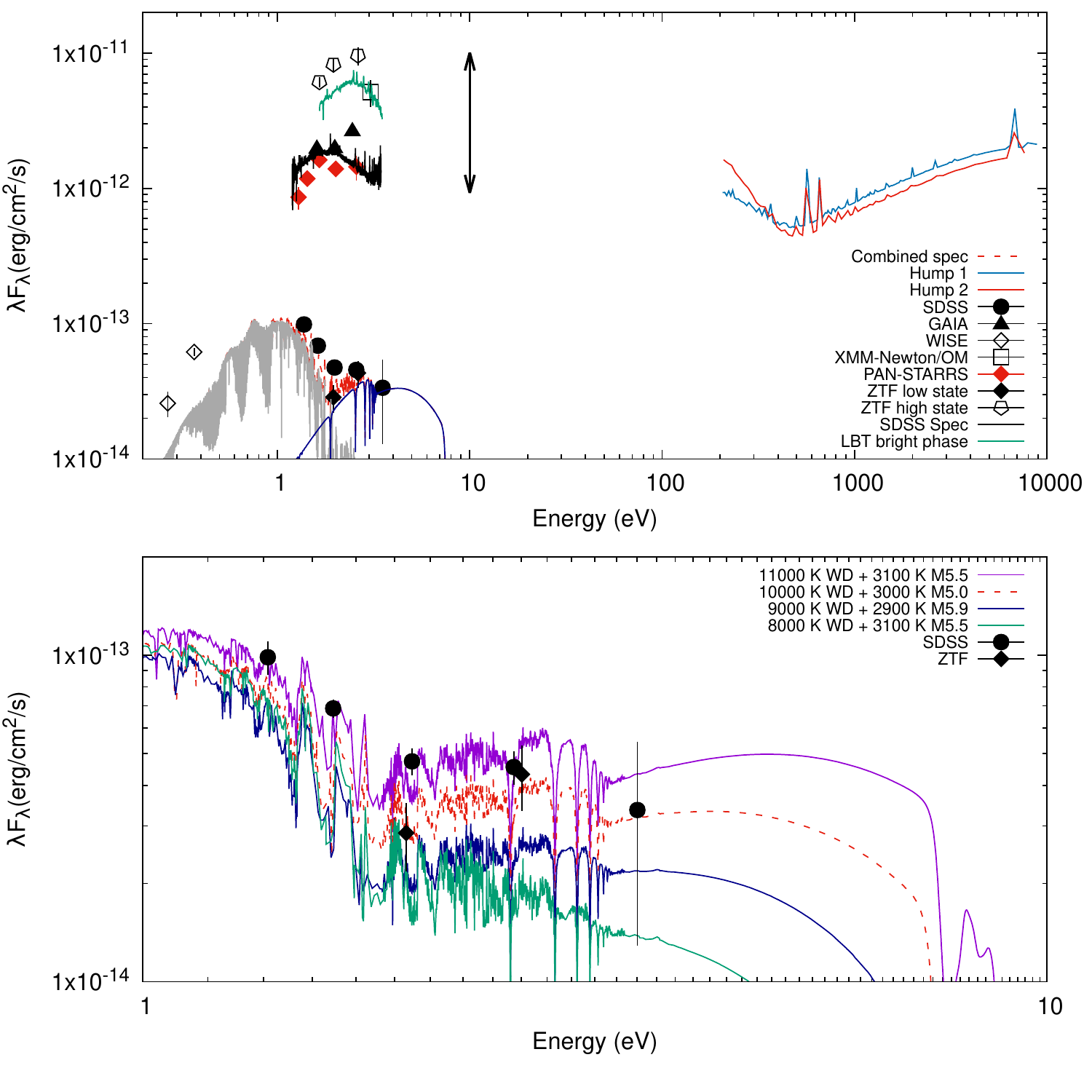}
  \caption{Upper panel: Total spectral energy distribution of \vum from the X-ray to the infrared region. White dwarf (0.8 $M_{\odot}$) (dark blue line) and main-sequence star (0.16 $M_{\odot}$) (light gray line) model spectra for 10000 K and 3000 K, are shown combined with a red dashed line. The vertically plotted black arrow represents the amplitude of the high state light curve in ZTF \textit{g}-band (16 - 18.6 mag). Lower panel: This graph focuses on the points in the optical region and combined synthetic atmosphere models that we use to estimate the temperatures of the components. }
      \label{f:sed}
  \end{figure}

\subsection{Spectral Energy Distribution \label{s:sed}}

The spectral energy distribution (SED) of \vum which was compiled using data from \xmmn, the Large Binocular Telescope (LBT) \citep{littlefield+18}, the Zwicky Transient Facility \citep[ZTF;][]{masci+19} (see Fig.~\ref{f:ztf}), the Sloan Digital Survey \citep[SDSS;][]{blanton+17}, the Panoramic Survey Telescope and Rapid Response System \citep[Pan-STARRS;][]{chambers+16} survey and the Wide-field Infrared Survey \citep[WISE;][]{wise+10} using effective wavelengths and zero points for each filter as given in the Spanish Virtual Observatory\footnote{\protect\url{http://svo2.cab.inta-csic.es/theory/fps/}} (SVO). For the WISE observations, we used bands W1 and W2 only, because the W3 and W4 band data are regarded as unreliable for targets as faint as \vum \citep{wise+10}. Fig.~\ref{f:sed} shows the spectral energy distribution generated from these observation points.
 
The various spectroscopic and photometric observations obtained at different times reveal two fundamentally different mass accretion states, a high and a low state, as seen in many other polars. The data analyzed by us from \tes and \xmmn can be understood in terms of a high state, but the archives (SDSS, ZTF) also document a low state. The high-state brightness at a given phase, e.g.~the peak brightness of the first hump, does not seem constant, but we do not attempt to subcategorize the state of enhanced activity further.

In Fig.\ref{f:sed}, we show one single data point representing the first hump's brightness. The white-light filter of \xmmn/OM covers a broad wavelength range, $\sim 200 - 8000$\,\AA. We use the central wavelength and flux conversion factors as given in the \xmmn online documentation\footnote{\protect\url{https://xmm-tools.cosmos.esa.int/external/xmm_user_support/documentation/uhb/omfilters.html}} to convert count rates to physical fluxes. The OM data seems to stay a bit below the brightest fluxes from the ZTF but still indicates a high state.

\subsubsection{Low accretion state \label{s:low}}

The long-term light curves from the ZTF (Fig.~\ref{f:ztf}) demonstrate different mass accretion states for \vum. The object switched from a high to a low state in May 2020. This transition was observed in the \textit{g} and \textit{r} filters, the ZTF database has no $i$ band data point after May 2020. The faintest points from the ZTF and data points from SDSS were used in the SED and are assumed to represent upper limits for the temperatures of the naked WD and the donor star. Thus, these points became a reference for us when estimating the temperatures of the components. 

We used the data of Koester \citep{koester+10,trembley+10} and the NextGen \citep{hauschildt+99} synthetic atmosphere models that are available online\footnote{Synthetic atmosphere data available at \protect\url{http://svo2.cab.inta-csic.es/theory/newov2}}, to represent the photospheric contributions of the WD and the donor star, which we assume to be a main-sequence star. To do this, we scale these models to the \gai-distance of 758 pc. The distance error causes uncertainty of 200 K in the temperatures of the WD and the donor at fixed masses. 

In the SED, we considered two main reference points, the ZTF $r$-band and the SDSS $u$-band. The $r$-band data point has contributions from both photospheres, while the contribution of the donor star to the SDSS $u$-band is negligible. The temperature of the white dwarf is constrained by this data point. An upper temperature for the donor is derived by requesting that the summed light of both model components shall not exceed the measured $r$-band brightness. 

To create the synthetic flux distribution, values for the free parameters, the radius $R_{\rm wd}$ and the effective temperature, \teff, need to be chosen. We used an 0.8 $M_{\odot}$ mass WD, which is assumed to be representative of magnetic CVs \citep[][and references therein]{pala+21}, and estimate $R_{\rm wd}$ using the \citet{nauenberg+72} mass-radius relation. The best matching temperature that we found to represent SED for this mass was 10000 K. Taking into account the error in the $u$ band magnitude, we estimate the WD temperature to be in the range 8000 and 11000 K (see Fig.~\ref{f:sed}). For the secondary star, we used Knigge's empirical relationship for CVs \citep{knigge+11}. An orbital period of 91.05 min suggests a secondary star with spectral type M6.5 (2750 K). A corresponding photospheric model underpredicts the observed optical data, which are better represented with a higher temperature of the donor (see Fig.~\ref{f:sed}), but we cannot exclude that some residual non-photospheric emission is present (cyclotron radiation, stream emission). 
In conclusion, the most consistent temperatures with optical data points were 10000$^{+1000}_{-2000}$ and 3000$^{+100}_{-250}$ K for the WD and companion, respectively. 
The best-fitted model is given in the lower panel of Figure~\ref{f:sed} as 10000 K WD + 3000 K M5 and is shown with a red dashed line.  On a side note, we would like to emphasize that these temperatures represent the upper limits for the components.

The WISE W1 and W2 band fluxes were too high to be compatible with the synthetic stellar models, which could mean that when the WISE data were collected, \vum was in a high state and these points probably have cyclotron emission contribution. Because of that, these data points are omitted here.

\subsubsection{High Accretion State \label{s:high}}

In the ZTF high-state data, i.e.~in all data obtained before May 2020, \vum varies between 16 and 18.6 in all three passbands. In Fig.~\ref{f:sed} we show the maximum brightness in each of the bands, which always corresponds to the first hump, and indicate the photometric variability with a vertical arrow.

The \gai and \textit{Pan-STARRS} data points in the Figure are averages over many individual measurements and fall within the range of the ZTF data. Hence, we assume that all these data were obtained during the high state. 

The \tes observations coincide with the time frame of the ZTF observations. Fig.~\ref{f:ztf} shows this interval in \textit{g}-band observations when the source varied between 16.83 - 17.65, i.e. \tes observations were made in the high state.

In Fig.~\ref{f:sed} we also show two optical spectra, one obtained from the SDSS database, the other obtained by \cite{littlefield+18}. The mean SDSS-spectrum was obtained in 28-JAN-2018.
The mean spectral flux density is about 3.0$\times$10$^{-16}$ erg cm$^{-2}$ s$^{-1}$ \mbox{\,\AA}$^{-1}$ at 5500\,\AA\, which corresponds to a V-band magnitude of 17.7 (Vega). The implied $g$-band magnitude is 17.9 (AB). The brightness is at the lower end but still compatible with a high state. Because the data were obtained shortly before ZTF started observing, we have no independent confirmation for its accretion state, but the rich emission lines spectrum indicates a high state.

\begin{figure}
   \centering
   \includegraphics[width=\columnwidth]{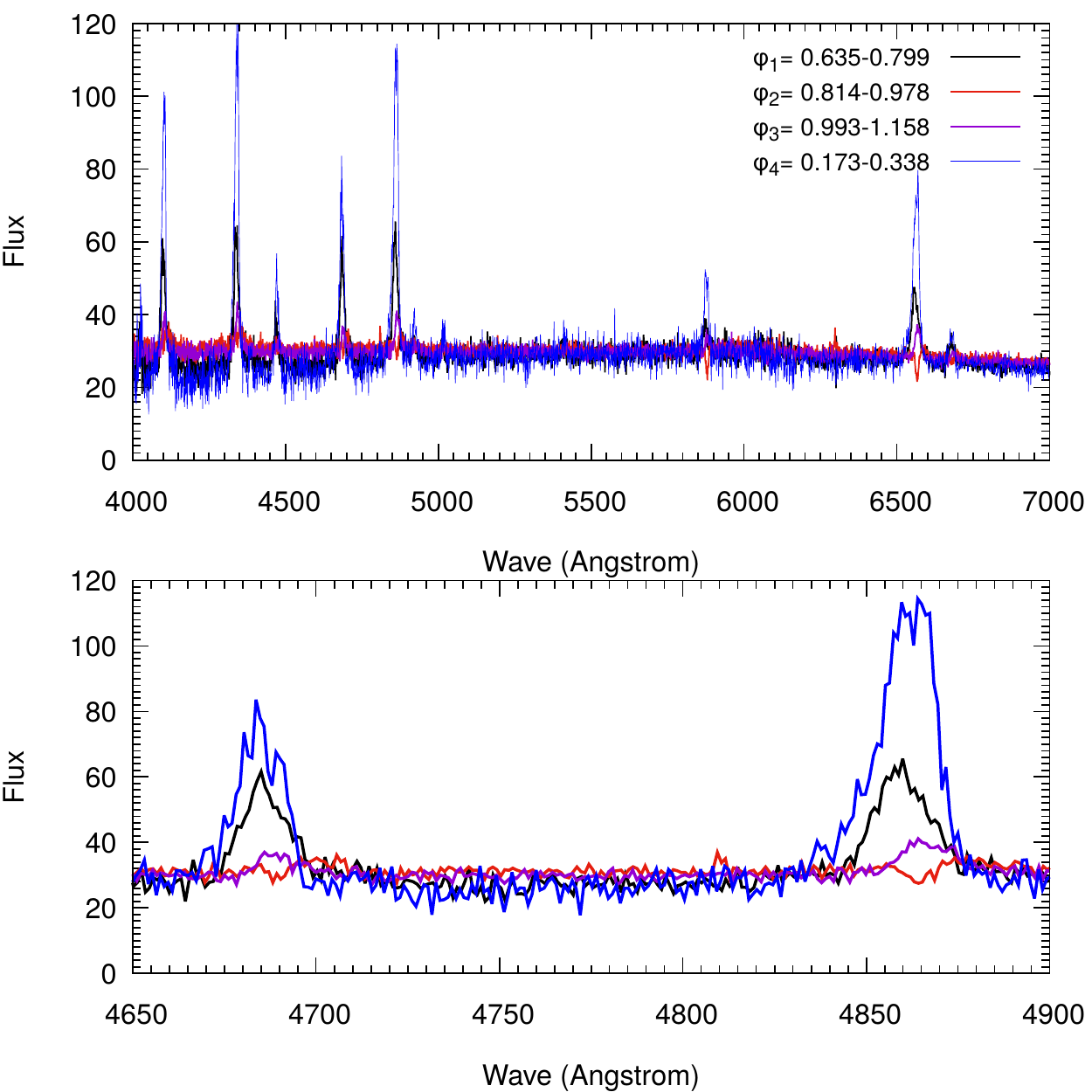}
      \caption{The upper panel shows four SDSS spectra which covered between $\phi_{\rm 1-4}$=0.635 - 1.338 phase intervals. The lower panel shows the wavelength region, which covers the H$\beta$ and the He{\sc II} 4686 emission lines. Flux units are 10$^{-17}$ erg cm$^{-2}$ s$^{-1}$ \mbox{\,\AA}$^{-1}$.}
      \label{f:sdss}
  \end{figure}

The mean SDSS spectrum consists of four spectra, each with 900 sec exposure time. In Fig.~\ref{f:sdss}, these are shown separately. We calculated the phase intervals covered by the four spectra with Eq.~1. The phase intervals are $\phi_{\rm 1} = 0.635 - 0.799$, $\phi_{\rm 2} = 0.814 - 0.978$, $\phi_{\rm 3} = 0.993 - 1.158$, and $\phi_{\rm 4} = 0.173 - 0.338$, respectively. Hence, the spectra cover the phase interval showing the decrease from the second hump, the dip phase, and the rise toward the first hump. Interestingly, there is not much variability in the continuum but a lot in the lines. The second spectrum, which covers the dip, displays the emission/absorption line reversal reported earlier by \citet{littlefield+18}. The composition of the emission line spectrum and the occurrence of the emission/absorption line reversal during the dip phase confirm that these data were obtained in a high accretion state. The SDSS spectra were obtained in an active state although not at the highest possible level. This becomes evident from the comparison of the two spectra in Fig.~\ref{f:sed} which shows the SDSS spectrum clearly below the LBT spectrum. To assess the energy balance in the high state, we therefore use only the LBT data.

The LBT-spectroscopy presented by \citep{littlefield+18} was obtained in a high state and was used to determine the cyclotron luminosity in the high state. Their simultaneous photometry showed higher peak brightness than the reduced spectrum indicating some light loss at the spectrograph's slit (slit width 1 arcsec in up to 1.3 arcsec ambient seeing, see their Fig.~4). We thus correct the observed spectra by a factor of 2. The spectrum at phase $\phi=0.99$ suggests that the cyclotron component peaks at around 6000\,\AA, and falls off to the IR wavelength regime.

Following \cite{littlefield+18} the difference spectrum between bright (first bright hump) and faint phase was regarded as the cyclotron spectrum proper and integrated. One may assume to have sampled a good fraction of the total cyclotron spectrum, which stretches further into the unobserved neighboring wavelength regions. By integrating the spectrum, we calculated the cyclotron flux as a $F_{\rm cyc}\sim 2.8\times10^{-12}$ ergs cm$^{2}$ s$^{-1}$ with a considerable but difficult to specify uncertainty. The $F_{\rm cyc}$ flux that we have obtained here is almost a factor 2 above the blackbody flux $F_{\rm bb}$ observed with \xmmn.

The cyclotron luminosity and the X-ray luminosity are accretion-induced, and their sum is regarded as a fair estimate of the accretion luminosity ($L_{\rm acc} \sim L_{\rm x} + L_{\rm cyc}$). However, one needs to recognize that the cyclotron luminosity is poorly constrained.
Uncertainties arise from the unknown bolometric correction factor for the flux and the unknown geometry factor to convert the flux to a luminosity. We assume $L_{\rm cyc}=\pi F_{\rm cyc}D^{2}$ to get $L_{\rm cyc}=4.8(2)\times10^{31}$ erg s$^{-1}$) at a distance of 758 pc.  The bolometric X-ray luminosity which we obtained from the first hump is $L_{\rm x}=2.3(3)\times10^{32}$ erg s$^{-1}$ so that the accretion luminosity from the first hump becomes $L_{\rm acc}=7.1(5)\times10^{32}$ erg s$^{-1}$.

We found the mass accretion rate ($\dot{M}=L_{\rm acc}R_{\rm wd}/M_{\rm wd}G$) in the high state of the \vum for the first bright hump as $\dot{M}_{\rm hump1}=$ $7.4(3)\times10^{-12}$ $M_{\odot}$ yr$^{-1}$. We used the same method for the second hump. The X-ray luminosity for second hump is $L_{\rm x}=2.5(4)\times10^{32}$ erg s$^{-1}$ and the estimated cyclotron flux $L_{\rm cyc}=2.9(2)\times10^{31}$ erg s$^{-1}$ which estimated from scaling the brightness of the first and the second humps and assuming the same spectral shape. The mass accretion rate was estimated for the second hump as $\dot{M}_{\rm hump2}= 6.8(5)\times10^{-12}$ $M_{\odot}$ yr$^{-1}$. Finally, the total accretion rate at the time of the \xmmn observation was $\dot{M}_{\rm tot} = \dot{M}_{\rm hump2}+\dot{M}_{\rm hump1} = 1.4(8) \times 10^{-11}$ M$_{\odot}$ yr$^{-1}$.

\subsection{Accretion Geometry}

In the following, we discuss constraints on the accretion geometry, which means the possible locations of the accretion spots, and the orbital inclination. We base the analysis and discussion on the phase difference of the humps in the light curves, the visibility of the two regions (or their self-eclipses), the existence of an X-ray absorption dip, and the non-existence of a binary eclipse.  

The X-ray observations revealed two bright phases per photometric cycle, to which we assign two accretion regions. We determined their centroids by applying Gaussian fits to the original light curve data. The first hump (red regions in Fig.~\ref{f:xmmlc}, 5 data points) has centroids at phases $\phi_{\rm phot} = 0.949, 0.982, 0.956, 1.012,$ and 0.995, respectively. The first two of the second humps are centered on phases of 0.479 and 0.515 with a mean of 0.497 (blue regions in Fig.~\ref{f:xmmlc}). These hump centroids scatter around the mean centroid phase of the first hump (given by phase $\phi_{phot}=0.0$) between $-18\degr$ and $+4\degr$ and the second hump between $172\degr$ to $186\degr$. In sum, the two regions are separated by about 0.5 phase units. This is particularly interesting because both regions have comparable X-ray luminosities and possibly similar accretion rates. Since there is no eclipse in the light curve, we do not know their absolute phase zero. The same problem was described by \citet{littlefield+18} who argue, based on expectation and similarity with other polars, that the phase of their observed emission/absorption line reversal is expected at binary phase $\sim$0.9, when the line-of-sight crosses the accretion stream.

\begin{figure}
   \centering
   \includegraphics[width=\columnwidth]{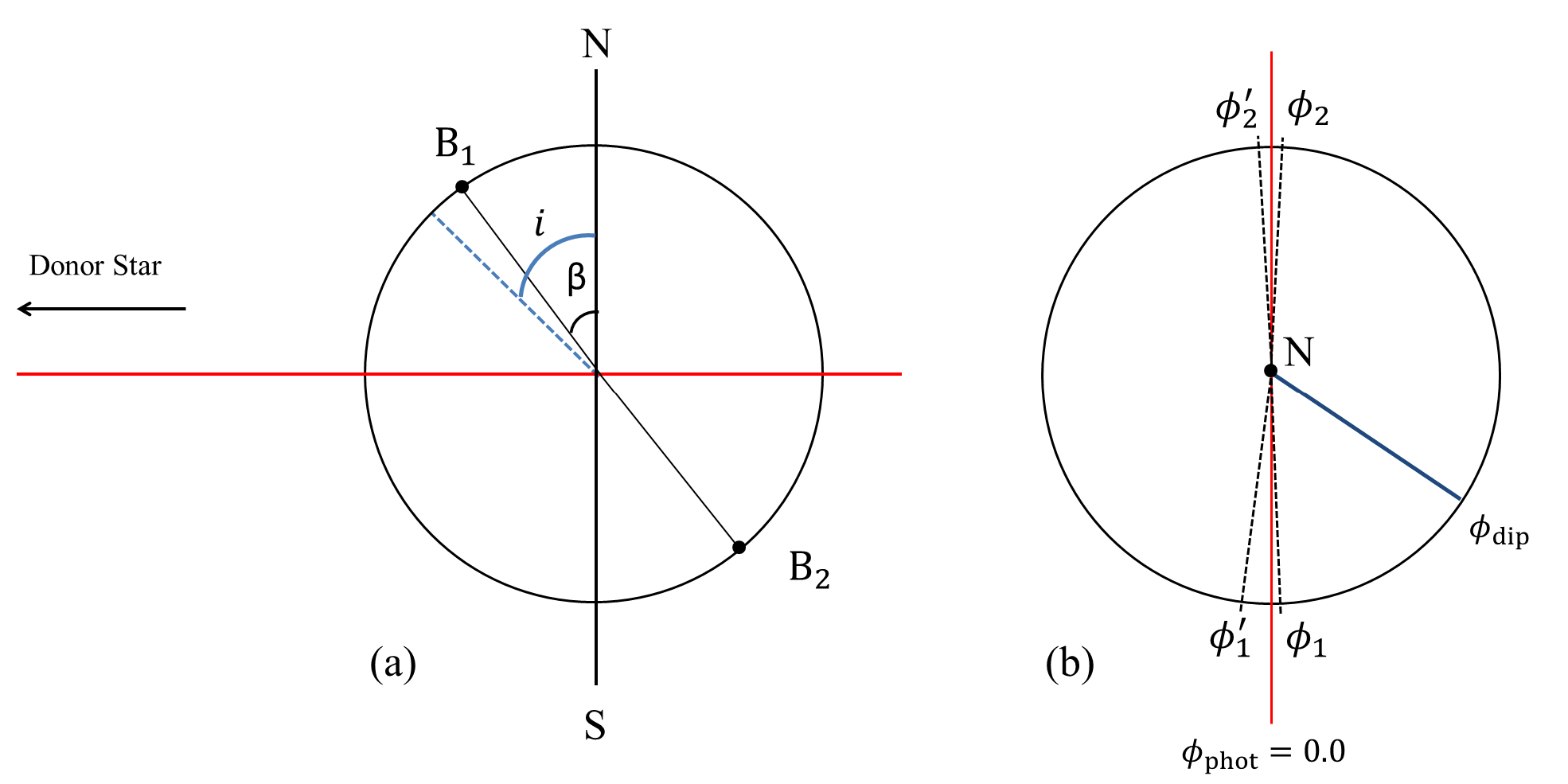}
      \caption{Schematic view of the two-pole accreting WD. The big black circles show the photosphere of the WD. (a) Side view of the WD. The red line represents the orbital plane. The dashed blue line shows the observer's line of sight. The magnetic accretion regions B1 (primary) and B2 (secondary) are almost diametrically opposed. (b) Top view, projection onto the orbital plane. $\phi_{phot}$ indicates photometric phase zero (center of the main hump). Phase is counted clockwise. $\phi_{1}$, $\phi^{'}_{1}$ and $\phi_{2}$,  $\phi^{'}_{2}$ represent the longitude range where the first and second column can be found on the WD surface in the \xmmn observations, respectively. The blue line indicates the phase, $\phi_{dip}$, of the X-ray absorption dip.}
      \label{f:geometry}
  \end{figure}

The X-ray data reveal an absorption dip at photometric phase $\phi_{\rm phot} = 0.81$. The existence of a dip requires $i> \beta$, with $i$ being the orbital inclination and $\beta$ the co-latitude of the corresponding accretion spot which is fed by this stream (see Fig.~\ref{f:geometry}). 

In other polars such a dip occurs between binary phases $\sim$0.93 and $\sim$0.8 (\citealp[V808 Aur, ][]{worpel+15}; \citealp[AM Her, ][]{schwope+20}). If this applies to \vum, we may expect the orbital phase of hump1 between phases 0.0 and 0.12. Hence, the spot faces the donor star directly or trailing the donor in phase by almost 45\degr. In any case, if the stream feeding the first region initially follows a ballistic trajectory in the WD's Roche lobe, it must be bent considerably so that the stream advances in the orbital phase. In such a geometry, the second region has its maximum at binary phase $\sim 0.5 - 0.6$. 

The X-ray light curve is modulated according to the visibility of the two accretion spots, which both seem to undergo self-eclipses by the white dwarf. It is difficult to determine the length of self-eclipse (or alternatively their visibility) individually because of apparent phases of visibility of both spots simultaneously. We roughly determined the visibility of the first hump when the second region was faint, i.e., after cycle 2.5 in Fig.~\ref{f:xmmlcclr}. We derive visibility of $\Delta\phi \simeq 0.55$, i.e. the first region lies above the orbital plane ('northern' spot). It thus appears likely that the X-ray absorption dip belongs to the stream feeding this region. However, we cannot safely determine the visibility of the second region, we assume nevertheless that is a southern region, i.e.~lies below the orbital plane. If this interpretation is correct, the dip lies at the beginning of the first bright phase, which is somewhat unusual. In other systems showing a dip like EK UMa \citep{clayton+94} and HU Aqr \citep{schwope+01}, the dip is located closer to the centers of the bright phase. A possible accretion geometry in projection onto the orbital plane ($xy$-plane) and a plane orthogonal to it ($xz$-plane) is sketched in  Fig.~\ref{f:geometry}.

\section{Discussion and conclusion \label{s:dis}}

This study investigates the thermal, temporal, and geometric behavior of \vum in the X-ray and optical regime. Unfortunately, the system does not show any eclipse. However, we have obtained the photometric period precisely from the extended \tes data connected without cycle count alias to data obtained with \xmmn obtained two years earlier. The improved period is based on the repetitive double hump structure.

There is a 3.8$\sigma$ difference between the period derived by \citet{littlefield+18} and our newly derived period. Precise (spectroscopic) observations are needed to understand better the reasons for this difference (a true period shift, an instrumental or a systematic error).

Also, the X-ray light curve is double-humped and shows a pronounced dip. Our analysis revealed that the dip at photometric phase $\phi_{\rm phot} = 0.81$ is shaped by photoelectric absorption due to the accretion stream crossing the line of sight. As a result, the second hump displays considerable brightness but no hardness variations. We interpret this behavior in a scenario with two different accretion poles. The brightness changes at the second pole are thus interpreted as accretion rate changes and, for example, not due to an intervening accretion curtain.

\citet{littlefield+18} described an emission-to-absorption line reversal which occurred at their photometric phase 0.87. We assume that the line reversal and the X-ray absorption dip are due to the same underlying effect, namely radiation interaction with the accretion stream which has left the orbital plane. We measured the X-ray absorption dip at our photometric phase of 0.81. Compared to \cite{littlefield+18} the phase difference between the dip and the center of the main hump changed from $\Delta\phi$=0.13 to 0.19. Such a variation is thought to be dependent on the mass accretion rate. The accreted matter may penetrate the magnetosphere deeper at high accretion rates, and the region where the material couples onto the magnetic field move away from the line connecting the WD and the donor star, causing the dip to occur at earlier orbital phases. A similar event was measured on eclipsing polar HU Aqr at different mass accretion states \citep{schwope+01}.

Using \xmmn, we determined the relative longitudes of the two accretion regions.
We assumed that the main accretion column (first hump) is in the direction of the secondary star or further ahead in phase by up to $\sim$0.1 phase units. Humps are mobile in the light curve, and the hump centers show some phase, hence azimuthal variability. 

The accretion geometry of \vum thus seems to be unusual and peculiar. The main spot seems to be ahead in phase. It is fed by a  stream that gives rise to the absorption dip, which must be bent profoundly to reach the main pole. The secondary pole might be similarly bright or even brighter in soft X-ray energies (see Fig.~\ref{f:xmmlc}, top panel) than the putative prime pole but is located in the hemisphere away from the donor star.

The previously reported cessation of mass accretion flow to the second column was seen in \tes and \xmmn observations. The weakening of the second column occurs randomly. We did not find any evidence that it followed a particular pattern. The \tes observations show that both poles seem to be equally affected by the change in the overall mass accretion rate. The \tes observations were obtained in a high accretion state but also included mass accretion cessations in some cycle intervals and their transitions. A general transition from a high to a low state was not observed with \tes. During these shirt-term transitions to a reduced state, no overall change in the shape of the light curve was detected. The second accretion pole also weakly represents itself in the \tes low state. Although the cases where the second pole is mainly weak in the low state, this situation has also been encountered in the high state. The location of the second pole, which is far from the secondary, and the partial mass accretion may be causing the interruption of the accretion to the second pole.

\vum exhibits a multi-temperature spectrum in X-rays. The system has a prominent blackbody emission at soft X-ray energy, which is typical for polars \citep{mukai17}. The black body ($kT_{\rm bb}$) temperature in polars spread between 5 - 30 eV \citep{kuulkers+10}. \vum has a hotter $kT_{\rm bb}$ ($\sim$40 eV) component than this for both poles and both models with high error values. Since these temperature values are outside the detection limits of \xmmn (< 0.2 keV), the peak temperature of the soft component comes with high error values. Nevertheless, lower limits in errors are still within this temperature range. 

The system shows two different mass accretion states in the spectral energy distribution, an active state at different brightness levels and an inactive or low state. \vum was in a high state during \xmmn observation. In a high state, the bolometric X-ray luminosity ($\approx5\times10^{32}$) lies upper energy limits which specified for polars \citep{verbunt+97, reis+13}. This value corresponds to the average X-ray luminosity of the Intermediate Polars \citep{patterson+94}. This luminosity is of particular interest given the short orbital period of the system (which would imply a lower accretion rate than systems with a longer orbital period) if our photometric period is the proper orbital period. Perhaps measuring such high luminosity may be due to the two-pole accretion. Other polars usually have one main pole with a typically lower field strength that is more or less facing the donor star and a second weakly accreting pole found at odd locations and often has the higher field strength (\citealp[UZ For, ][]{schwope97};  \citealp[AM Her, ][]{campbell+08a}; \citealp[VV Pup, ][]{campbell+08b}). However, the most significant property that makes \vum different from other polars is the almost equal contribution of the two emission regions to the overall energy budget. Considering the variable second hump, it is curious how the same amount of matter was carried to the second column. In the case of the self eclipse of columns, the co-latitude angle should be smaller than the inclination ($\beta < i$). The second column is located quite far from the stagnation region in such a geometry. In order for the matter to reach the second pole, it should be considered that the orbital phase zero must be in a later phase, likely $\phi_{phot}\sim0.1$. Thus, it can be ensured that a sufficient amount of mass can be conveyed from the stagnation region to both poles, and certain interruptions may occur on the second pole with the change in the mass accretion rate. Similar mass accretion rate variations onto a second pole were also observed in a similar system, V808 Aur \citep{worpel+15}.

We tried to constrain the parameters of the components that make up the system with synthetic spectra that we used in spectral energy distribution. Significantly, the reference of optical data points, with the distance from \textit{Gaia}, the SED tells us the upper-temperature limit of the white dwarf can be in the range of 11000 - 9000 K with the mass of 0.8 $M_{\odot}$. It would be possible to obtain similar SED distribution for different white dwarf masses at different temperatures. Since we do not have direct observational evidence to determine the mass of the WD, we used the currently accepted mean mass of WDs in polars.

Our attempt to find the spectral type of the secondary was based on the empirical relationship of \citet{knigge+11}. A main-sequence star of the M5.0 spectral type, together with a 0.8 $M_{\odot}$ white dwarf and the temperature of 10000 K, represented quite well the spectral energy distribution in the low accretion state. 

The parameters of the WD in \vum would benefit from a dedicated spectroscopic UV observation in the low state. This is because the contribution of the donor is negligible in the UV. Hence the temperature of the WD and its radius would be much better constrained if the WD could be unequivocally be identified and traced spectroscopically in the UV.

Finally, we emphasize the importance of spectroscopic observations to identify features from the secondary star (either in emission or absorption) to constrain the orbital motion of the secondary and orbital phase zero. This would be particularly useful in the low state, where the photospheric absorption lines should be easier to detect.

---------------------------------------------
\begin{acknowledgements}
This research has made use of data, software and/or web tools obtained from the High Energy Astrophysics Science Archive Research Center (HEASARC), a service of the Astrophysics Science Division at NASA/GSFC and of the Smithsonian Astrophysical Observatory’s High Energy Astrophysics Division. This paper includes data collected by the \tes mission. Funding for the \tes mission is provided by the NASA's Science Mission Directorate. Funding for the Sloan Digital Sky 
Survey IV has been provided by the 
Alfred P. Sloan Foundation, the U.S. 
Department of Energy Office of 
Science, and the Participating 
Institutions. 

SDSS-IV acknowledges support and 
resources from the Center for High 
Performance Computing  at the 
University of Utah. The SDSS 
website is www.sdss.org.

SDSS-IV is managed by the 
Astrophysical Research Consortium 
for the Participating Institutions 
of the SDSS Collaboration including 
the Brazilian Participation Group, 
the Carnegie Institution for Science, 
Carnegie Mellon University, Center for 
Astrophysics | Harvard \& 
Smithsonian, the Chilean Participation 
Group, the French Participation Group, 
Instituto de Astrof\'isica de 
Canarias, The Johns Hopkins 
University, Kavli Institute for the 
Physics and Mathematics of the 
Universe (IPMU) / University of 
Tokyo, the Korean Participation Group, 
Lawrence Berkeley National Laboratory, 
Leibniz Institut f\"ur Astrophysik 
Potsdam (AIP),  Max-Planck-Institut 
f\"ur Astronomie (MPIA Heidelberg), 
Max-Planck-Institut f\"ur 
Astrophysik (MPA Garching), 
Max-Planck-Institut f\"ur 
Extraterrestrische Physik (MPE), 
National Astronomical Observatories of 
China, New Mexico State University, 
New York University, University of 
Notre Dame, Observat\'ario 
Nacional / MCTI, The Ohio State 
University, Pennsylvania State 
University, Shanghai 
Astronomical Observatory, United 
Kingdom Participation Group, 
Universidad Nacional Aut\'onoma 
de M\'exico, University of Arizona, 
University of Colorado Boulder, 
University of Oxford, University of 
Portsmouth, University of Utah, 
University of Virginia, University 
of Washington, University of 
Wisconsin, Vanderbilt University, 
and Yale University.
Samet Ok is supported by TUBITAK 2219-International Postdoctoral Research Fellowship Program for Turkish Citizens. We are grateful to the anonymous referee, whose comments led to large improvements in the clarity of the paper.
\end{acknowledgements}

%
%

\bibliography{vum}
\bibliographystyle{aa}

\end{document}